\documentclass[12pt,a4paper]{article}
\usepackage{amssymb}
\usepackage{amsfonts}
\usepackage{amsmath}

\newcommand*{\cck}
[1]{\stackrel{*}{#1}{\vphantom{#1}}}
\newcommand*{\cc}
[1]{ \rlap{$\stackrel{*}{\phantom{#1}}$}#1 }

\newcommand{\mur}{{\underline{m}}}
\newcommand{\nur}{{\underline{n}}}

\newcommand{\alphar}{{\underline{a}}}
\newcommand{\betar}{{\underline{b}}}
\newcommand{\da}{{\dot\alpha}}

\newcommand{\sign}{\mathop{{\rm sign}}}

\newcommand*{\ccc}
[1]{\stackrel{*}{#1}\!\!{\vphantom{#1}}}

\newcommand{\uz}{{\underline z}{\vphantom{z}}}

\newcommand{\barM}{\hat{\bar M}{\vphantom{M}}}
\newcommand{\barMM}{\hat{\bar{\mathbf{M}}}{\vphantom{M}}^2}
\usepackage[left=2.7cm, right=2.2cm, top=2cm, bottom=2cm]{geometry}
\begin{document}

\title{Group-theoretical classification of orientable objects and particle
phenomenology}
\author{D. M. Gitman$^{1,2}$\thanks{%
dmitrygitman@hotmail.com} and A. L. Shelepin$^{3}$\thanks{%
alex@shelepin.msk.ru} \\
$^{1}$ P.N. Lebedev Physical Institute, \\
53 Leninskiy ave., 119991 Moscow, Russia.\\
$^{2}$ Institute of Physics, University of S\~{a}o Paulo, \\
Rua do Mat\~{a}o, 1371, CEP 05508-090, S\~{a}o Paulo,, Brazil\\
$^{3}${MIREA -- Russian Technological University,}\\
Prospect Vernadskogo, 78, 117454, Moscow, Russia }
\maketitle

\begin{abstract}
In our previous works, we have proposed a quantum description of relativistic orientable objects by a scalar field on the Poincar\'{e} group. This description is, in a sense, a generalization of ideas used by Wigner, Casimir and Eckart back in the 1930's in constructing a non-relativistic theory of a rigid rotator. The present work is a continuation and development of the above mentioned our works.  The position of the relativistic orientable object in Minkowski space is completely determined by the position of a body-fixed reference frame with respect to the space-fixed reference frame, and can be specified by elements $q$ of the motion group of the Minkowski space - the Poincar\'e group $M(3,1)$. Quantum states of relativistic orientable objects are described by scalar wave functions $f(q)$ where the arguments $q=(x,z)$ consist of Minkowski space-time points $x$, and of orientation variables $z$ given by elements of the matrix $Z\in SL(2,C)$. Technically, we introduce and study the so-called double-sided representation $\boldsymbol{T}(\boldsymbol{g})f(q)=f(g_l^{-1}qg_r)$, $\boldsymbol{g}=(g_l,g_r)\in \boldsymbol{M}$, of the group $\boldsymbol{M}$, in the space of the scalar functions $f(q)$. Here the left multiplication by $g_l^{-1}$ corresponds to a change of space-fixed reference frame, whereas the right multiplication by $g_r$ corresponds to a change of body-fixed reference frame. On  this basis, we develop a classification of the orientable objects and draw the attention to a possibility of connecting these results with the particle phenomenology. In particular, we demonstrate how one may identify fields described by linear and quadratic functions of $z$ with known elementary particles of spins $0$,$\frac{1}{2}$, and $1$. The developed classification does not contradict the phenomenology of elementary particles and, moreover, in some cases give its group-theoretic explanation.
\end{abstract}

\section{Introduction\label{S1}}

It is known that the first and most known example of a theory in which wave
functions depend on some orientation variables is the non-relativistic
theory of a rigid rotator, constructed by Wigner, Casimir and Eckart back in
the 1930's (see Ref. \cite{BieLo81} for references and historical remarks).
In their construction, one reference frame (laboratory, space-fixed, \emph{%
s.r.f.} in what follows) is connected with surrounding objects, while
another one (body-fixed, \emph{b.r.f.} in what follows) is connected with a
rotating body. Correspondingly, there appear two sets of operators of the
angular momentum: one related to the \emph{s.r.f.} (left generators of the
rotation group $\hat{J}_{k}^{L}$), and another related to the \emph{b.r.f.}
(right generators of the rotation group $\hat{J}_{k}^{R}$). These two sets
of generators commute with each other, and the rotator Hamiltonian is
constructed from the right generators $\hat{J}_{k}^{R}$. The interpretation
of the generators $\hat{J}_{k}^{R}$ as projections of the momentum on the 
\emph{b.r.f.} belongs to Wigner and Casimir (1931), and underlies the theory
of molecular spectra. In fact, the the non-relativistic theory of the rigid
rotator is formulated as a theory of a scalar filed on the group $%
SO(3)\thicksim SU(2)$.

It should be also noted that a description of relativistic spinning
particles can be reformulated in terms of one scalar field depending on the
Minkowski space-time coordinates as well as on some auxiliary continuous
variables that describe spin. Such a reformulation has a long history,
starting with the work \cite{GinTa47} of Ginzburg and Tamm. At the end of
1940's and the beginning of\ 1950's, independently by various authors, see
Refs. \cite{GinTa47,BarWi48,Yukaw50,Shiro51} (mainly in connection
constructing of relativistic wave equations (RWE)), were introduced some
fields depending not only on the Minkowski space-time coordinates but also
on a certain set of spinning variables. A systematic treatment of these
fields as fields on homogeneous spaces of the Poincar\'{e} group was
presented by Finkelstein in 1955; see Ref. \cite{Finke55}. He also
considered a classification and explicit constructions of homogeneous spaces
of the Poincar\'{e} group that contains the Minkowski space which is a
homogeneous space of the latter group. In 1964, Lur\c{c}at suggested to
construct a quantum theory on the whole Poincar\'{e} group, instead on the
Minkowski space, see Ref. \cite{Lurca64}, but this idea did not receive
immediate development.

In 70-90s of the last century, ideas of considering fields on homogeneous
spaces of the Poincar\'{e} group, gained a certain development, in
particular, in the Refs \cite%
{BacKi69,Kihlb70,BoyFl74,Arodz76,Tolle78,Tolle96,Drech97,Hanni97,Varla04}.
Properties of different spaces were studied, as well as some possibilities
of introducing interactions in the spin phase-space and possibilities of
constructing corresponding Lagrangian formulations. It was found that the
choice of a homogeneous space generates restrictions on the corresponding
scalar fields. Thus, authors of the Ref. \cite{BacKi69} had arrived at the
conclusion that the minimal dimension of a homogeneous space suitable for a
description of integer and semi-integer spins is equal to eight.

In our previous works \cite{GitSh01,GitSh09,GitSh10,GitSh11}, following the
ideas of the pioneer works cited above, we have developed a quantum
description of relativistic orientable objects by a scalar field $f(q)$, $%
q=(x,z)$ that depends on the Minkowski space-time coordinates $x=\left(
x^{\mu }\right) =\left( x^{0},\mathbf{x}\right) $, $\mathbf{x=\ }\left(
x^{k}\right) ,$ and on the orientation variables $z=\left( z_{\alpha \beta
}\right) $ given by elements of the matrix $Z\in SL(2,C)$, in fact, on the
Poincar\'{e} group $M(3,1)$, $q\in M(3,1)$.

At the beginning, we have proposed an approach to constructing fields on
groups of motions in Euclidean and pseudo-Euclidean spaces, elaborating in
detail the cases of $2,3$- and $4$ dimensions; see Ref. \cite{GitSh01}. In
that construction, a scalar field appears as a generating function of usual
multicomponent fields. In particular, it has been demonstrated that, as
distinct from the case of scalar fields on homogeneous spaces, the field on a
group as a whole is closed with respect to discrete transformations. The
discrete transformations ($C,P,T$) correspond to involutive automorphisms of
the group and are reduced to a complex conjugation and to a change of the
arguments of scalar functions $f(q)$; see Ref. \cite{BucGiS02}. Most types
of RWE were written in terms of scalar functions $f(q)$.

By analogy with the non-relativistic theory of the rigid rotator, in the
relativistic case, it is supposed that one reference frame (laboratory,
space-fixed, \emph{s.r.f.}) is connected with surrounding objects, while
another one (body-fixed, \emph{b.r.f.}) is connected with an orientable
object. The position of the latter is completely determined by the position
of the corresponding \emph{b.r.f.} with respect to some \emph{s.r.f.}, such
that all the positions can be given by elements of the Poincar\'{e} group. 
It implies that quantum states of the relativistic orientable objects are 
described (as was already said above) by scalar wave functions $f(q)$ on 
the Poincar\'{e} group.

On this basis, in the Ref. \cite{GitSh09} two types of
transformations in the space of scalar functions $f(q)$ (see just below)
were defined and studied. In the Refs. \cite{GitSh10,GitSh11} a method of
classifying such objects with respect to the corresponding transformation
properties of the functions $f(q)$ was proposed, as well as possible ways of
introducing interactions of orientable objects. 

By the present consideration we develops results of the latter works. As
an important part of this development we represent a more complete
classification of the orientable objects and draw the attention of readers
to the possibility of connecting these results with the particle
phenomenology. In particular, we observe the possibility of identifying
fields described by linear and quadratic functions of $z$ with known
elementary particles of spin $0$,$\frac{1}{2}$, and $1$.

Now we pass to technical details of the mentioned above transformations. In
this regard, it is useful once again to stress that the position of an
orientable object is completely determined by the position of the
corresponding \emph{b.r.f.} with respect to some \emph{s.r.f.} . Symmetries
of the space containing the orientable object (space-time or external
symmetries) determine the behavior of wave functions with respect to the
left transformations $T_{L}\left( g_{l}\right) $, $g_{l}\in M(3,1)_{ext}$,
whereas the symmetries of the orientable object itself (internal symmetries)
determine the behavior of wave functions functions with respect to the right
transformations $g_{r}$, $g_{r}\in M(3,1)_{int}$, 
\begin{equation}
T_{L}\left( g_{l}\right) f(q)=f(g_{l}^{-1}q),\ \ T_{R}\left( g_{r}\right)
f(q)=f(qg_{r})\ .  \label{0}
\end{equation}
As it is known, operations (\ref{0}) define left and right regular
representations of the Poincar\'{e} group. Studying these representations is
equivalent to studying representation $\boldsymbol{T}(\boldsymbol{g})$, of
the group $\boldsymbol{M}=M(3,1)_{ext}\times M(3,1)_{int}$,\ $\boldsymbol{g}%
\in \boldsymbol{M},$ in the space of scalar functions $f(q),$%
\begin{equation}
\boldsymbol{T}(\boldsymbol{g})f(q)=f(g_{l}^{-1}qg_{r}),\ \ \boldsymbol{g}%
=(g_{l},g_{r})\in \boldsymbol{M}.  \label{1}
\end{equation}%
Here the left multiplication by $g_{l}^{-1}$ corresponds to a change of the 
\emph{s.r.f.} , whereas the right multiplication by $g_{r}$ corresponds to a
change of the \emph{b.r.f.} . It is easy to verify that Eq. (\ref{1})
specifies a representation $\boldsymbol{T}(\boldsymbol{g})$ of the group $%
\boldsymbol{M}$ in the space of the scalar functions $f(q)$. In what follows
we call such a representation the \emph{double-sided} representation.

A classification of functions $f(q)$ on a group is the subject of the
harmonic analysis on Lie groups and is carried out using the maximal sets of
commuting operators of the group; see e.g. \cite{BarRa77}. The number of the
operators in the maximal set is equal to the number of the group parameters,
and the set consists of the Casimir operators and of the (equal in number)
functions of the generators in left and right regular representations (left
and right generators).

It is important to emphasize that due to the fact that the groups $%
M(3,1)_{ext}$ and $M(3,1)_{int}$ act in the same space of function $f(q)$,
Casimir operators constructed from right and left generators coincide. This
means that states transformed by a fixed representation of $M(3,1)_{ext}$
are transformed by the same representation of $M(3,1)_{int}$.

In $3+1$ dimensions, according to the above said, and as a consequence of
Eq. (\ref{1}), there appear two sets of 10 generators, the right and the
left ones. Having in hands these generators one can construct a maximal set
of commuting operators acting in the space of the functions $f(q)$. The
maximal set of commuting operators consists of 10 operators -- two Casimir
operators, four functions of left generators and four functions of right
generators. The last four functions of generators $M(3,1)_{int}$, which are
not expressed in terms of generators $M(3,1)_{ext}$, commute with all the
generators $M(3,1)_{ext}$, i.e., they correspond to the internal quantum
numbers of the orientable object. The question arises about their physical
meaning and (due to the coincidence of the Casimir operators) the possible
connection of their eigenvalues with the eigenvalues of the generators of
external symmetries $M(3,1)_{int}$.

Thus, considering the two-sided representation of the group $\boldsymbol{M}%
=M(3,1)_{ext}\times M(3,1)_{int}$, we naturally come to the question of the
relation between external and internal symmetries and corresponding quantum
numbers. In this regards, one ought to mention that in the 1960s, attempts
were made to unite internal and external symmetries in the framework of a
one group. Soon, however, the so-called no-go theorem \cite{ColMa67} was
proved (under some very general assumptions), stating that the symmetry
group of the $S$-matrix is locally isomorphic to a direct product of the
Poincar\'{e} group and the group of internal symmetries. However, on this
basis, one sometimes makes too strong conclusion that a non-trivial relation
between internal and external symmetries is impossible.

As was already said, transformations (\ref{1}) form the direct product of
the groups of internal and external symmetries, which removes the case under
consideration from the prohibitions of this theorem. Moreover, as is
demonstrated below, a non-trivial relation between internal and external
quantum numbers is possible. Both transformation groups, corresponding to
changing the \emph{s.r.f.} and the \emph{b.r.f.}, are acting in the same
space of scalar functions $f(q)$ and have the same Casimir operators which
define the mass and the spin. By fixing eigenvalues of the Casimir
operators, and therefore fixing a representation, we obviously impose some
conditions on the spectra of both left and right operators that enter the
maximal set. Thus, in spite of the fact that the left and the right
generators commute, their spectra are not independent. Below we are observed
such a relationship on some concrete examples.

We note that the maximal sets of commuting operators are different for the
massless and the massive cases, that is why, in what follows, we consider
them separately.

By orientable objects, we understand states that are
transformed according to irrep of the group $M(3,1)_{ext}$. By fixing irrep
of $M(3,1)_{ext}$, we, as a consequence of the law of transformation (\ref{1}%
), fix irrep of $\boldsymbol{M}=M(3,1)_{ext}\times M(3,1)_{int}$. This means
that irreps (describing multiplets of particles) of $\boldsymbol{M}%
=M(3,1)_{ext}\times M(3,1)_{int}$ decompose into a sum of $M(3,1)_{ext}$
irreps, differing by eigenvalues of generators $M(3,1)_{int}$ (internal
quantum numbers). In other words, different particles are states belonging
to irrep $\boldsymbol{M}$, not connected by transformations of $M(3,1)_{ext}$%
. Note that, unlike the external symmetries the internal ones can be
violated, which is quite similar to the situation observed in the case of
the rigid rotator, for which internal symmetries are the symmetries of the
rotating body itself.

The work is organized as follows. In Sections \ref{S2} and \ref{S3}, we
construct and study maximal sets of commuting operators in the space of
functions $f(q)$ on the Poincar\'{e} group for the cases of the mass $m=0$
and $m\neq 0$. In Sections \ref{S4} and \ref{S5}, we represent a
classification of massless orientable objects of chirality $0,\pm 1/2,\pm
1,\pm 3/2$ and formulate corresponding RWE. In Section \ref{S6} we give a
classification of orientable objects with mass $m\neq 0$ and spin $0,1/2,1,3/2$%
. In Sections \ref{S7} and \ref{S8} we consider the ultra-relativistic limit
and formulate corresponding RWE.  A comparison of the obtained
classification of the orientable objects with the particle phenomenology is
given in Sect. \ref{S9}; some concluding remarks are summarized in the same
Section. In the Appendix (Section \ref{S10}) we construct generators of the
Lorentz group in the space of the functions $f\left( z\right) $.

\section{Field on the Poincar\'{e} group and maximal sets of commuting
operators \label{S2}}

The functions $f(q)$ on the Poincar\'{e} group $M(3,1)$ depend on $10$
parameters, 
\begin{equation}
q={X,Z},\quad X=\{x^{\mu }\},\quad Z= \{ z^\alpha_{\;\;\betar} \} = \left( 
\begin{array}{cc}
z_{\;\,\underline{1}}^{1} & z_{\;\,\underline{2}}^{1} \\ 
z_{\;\,\underline{1}}^{2} & z_{\;\,\underline{2}}^{2}%
\end{array}%
\right) =\left( 
\begin{array}{cc}
z^{1} & {\underline{z}}^{1} \\ 
z^{2} & {\underline{z}}^{2}%
\end{array}%
\right) \in SL(2,C),  \label{6}
\end{equation}
where $\det Z=1$.
Action of the groups $M(3,1)_{ext}$ and $M(3,1)_{int}$ in the space of functions $f(q)$ 
\begin{eqnarray}
&&M(3,1)_{ext}:\qquad T_{L}(g_l)f(q)=f(g_l^{-1}q),  \label{2} \\
&&M(3,1)_{int}:\qquad T_{R}(g_r)f(q)=f(qg_r),  \label{3} 
\end{eqnarray}
corresponds to the replacements of s.r.f. and b.r.f., respectively.
Representations (\ref{2}), (\ref{3}) provide the most general consideration, since regular
representations contain all (up to equivalence) irreducible representations (irreps) of the group \cite{Vilen68t}.
Transformations from $M(3,1)_{ext}$ correspond to the symmetries of the containing space, or external ones, 
and transformations from $M(3,1)_{int}$ corresponds to  symmetries of the orientable object, or internal symmetries.

The actions on the left and the right correspond to two sets of generators
of the Poincar\'{e} group. There are ten generators each in the left and 
right regular representations, which for brevity we will call the left and right
generators.

The generators that correspond to translations and rotations have the form 
\begin{eqnarray}
\label{gen.L}
&&M(3,1)_{ext}:\qquad\hat{p}_{\mu }=i\partial /\partial x^{\mu }, \quad
 \hat{J}_{\mu\nu }= \hat{L}_{\mu\nu }+ \hat{S}_{\mu \nu},
 \quad \mu,\nu =0,1,2,3, 
\\  \label{gen.R}
&&M(3,1)_{int}:\qquad\hat{p}_{\mur }^R=-v^\nu_{\;\;\mur} \hat p_\nu, \quad
\hat{J}_{\mur\nur }^R=\hat{S}_{\mur\nur }^R,
 \quad \mur,\nur =0,1,2,3,
\end{eqnarray}%
where $\hat{L}_{\mu \nu }=i(x_{\mu }\partial _{\nu }-x_{\nu }\partial _{\mu
})$ are the operators of orbital momentum projections and $\hat{S}_{\mu \nu
} $ are the operators of spin projections. Here 
$v^{\mu}_{\;\;\nur} = \frac{1}{2}(\sigma^\mu)_{\dot{\beta}\alpha}(\bar\sigma_\nur)^{\alphar\dot{\betar}}
z^{\alpha}_{\;\;\alphar}\cc z^{\dot\beta}_{\;\;\dot{\betar}}$,
the operators $\hat{S}_{\mu\nu}$ and $\hat{S}^R_{\mur\nur }$
depend only on $z$ and $\partial/\partial z$. Their explicit form
is given in the Appendix. It is also convenient to use a three-dimensional notations: 
$\hat S^k =\frac 12\epsilon_{ijk} \hat S^{ij}$, $\hat B^k =\hat S^{0k}$.
Generators from different sets commute with each other. 

In consequence of the unimodularity of $2\times 2$ matrices $Z$ there exist
invariant antisymmetric tensors $\varepsilon ^{\alpha \beta}=-\varepsilon
^{\beta \alpha }$, $\varepsilon ^{{\dot{\alpha}}{\dot{\beta}}}=-\varepsilon
^{{\dot{\beta}}{\dot{\alpha}}}$, $\varepsilon ^{12}=\varepsilon ^{{\dot{1}}{%
\dot{2}}}=1$, $\varepsilon _{12}=\varepsilon _{{\dot{1}}{\dot{2}}}=-1$. Now
spinor indices are lowered and raised according to the rules 
\begin{equation}
z_{\alpha }=\varepsilon _{\alpha \beta }z^{\beta },\quad z^{\alpha
}=\varepsilon ^{\alpha \beta }z_{\beta }.  \label{tens.5}
\end{equation}

Discrete transformations correspond to the automorphisms of the Poincar\'{e}
group, their action is reduced to replacing arguments and complex
conjugation of functions $f(x,z)$ \cite{BucGiS02,GitSh09}.

Classification of functions on the group is carried out using maximal sets
of commuting operators. A maximal set of commuting operators in the space of
the functions on the group contains Casimir operators and an equal number
operators that are some functions of left and right generators \cite{BarRa77}%
. In total, in our case, the set contains $10$ operators, according to the
number of group parameters.

Casimir operators, that label irreps, can be composed of both left and right
generators, so that the \textquotedblleft left\textquotedblright\ and
\textquotedblleft right\textquotedblright\ mass and spin are the same. For
the Poincar\'{e} group $M(3,1)$, we have: 
\begin{eqnarray}
&&\hat p^2=\eta^{\mu\nu}\hat p_\mu \hat p_\nu = \eta^{\mur\nur}\hat p^R_\mur \hat p^R_\nur,
\label{7}
\\
&&
 \hat W^2 =\eta^{\mu\nu}\hat W_\mu \hat W_\nu=\eta^{\mur\nur}\hat W^R_\mur \hat W^R_\nur, \quad
\label{8}
\\
&&
 \hbox{where} \quad
 \hat W_\mu=\frac 12\epsilon_{\mu\nu\rho\sigma}\hat p^\nu \hat J^{\rho\sigma}
 = \frac 12\epsilon_{\mu\nu\rho\sigma}\hat p^\nu \hat S^{\rho\sigma},\qquad
 \hat W^R_{\underline{m}} = \frac 12\epsilon_{{\underline{m}}{\underline{n}}\underline{r}\underline{s}}
 \hat p_R^{\underline{n}} \hat S_R^{\underline{r}\underline{s}}.
\label{W}
\end{eqnarray}
As four operators, composed of left generators (generators of the group 
$M(3,1)_{\mathrm{ext}}$), one can choose the Casimir operator 
$\hat{\mathbf{p}}\hat{\mathbf{J}}=\hat{\mathbf{p}}\hat{\mathbf{S}}$ and 
the generators $\hat{p}_{k}$ of the subgroup $M(3)_{\mathrm{ext}}$. 
The latter correspond to additive quantum numbers.
Eigenfunctions of these operators correspond to definite values of the
helicity and the momentum.

Functions of right generators can be chosen in different ways. We choose
sets, corresponding to reduction $M(3,1)_{\mathrm{int}}\supset SL(2,C)_{%
\mathrm{int}}$. Two Casimir operators of the \textquotedblleft
right\textquotedblright\ Lorentz group $SL(2,C)_{\mathrm{int}}$ have the
form 
\begin{equation}
\hat{\mathbf{S}}^{2}-\hat{\mathbf{B}}^{2}=\frac{1}{2}\hat{S}_{{\underline{m}}%
{\underline{n}}}^{R}\hat{S}_{R}^{{\underline{m}}{\underline{n}}}=\frac{1}{2}%
\hat{S}_{\mu \nu }\hat{S}^{\mu \nu },\quad \hat{\mathbf{S}}\hat{\mathbf{B}}=%
\frac{1}{16}\epsilon ^{{\underline{m}}{\underline{n}}\underline{r}\underline{%
s}}\hat{S}_{{\underline{m}}{\underline{n}}}^{R}\hat{S}_{\underline{r}%
\underline{s}}^{R}=\frac{1}{16}\epsilon ^{\mu \nu \rho \sigma }\hat{S}_{\mu
\nu }\hat{S}_{\rho \sigma }.  \label{10}
\end{equation}%
where $\hat{S}^{i}$ and $\hat{B}^{i}$ are operators of spin and boost
projections, see Appendix. In contrast to $\hat{S}_{{\underline{m}}{%
\underline{n}}}^{R}$, being the generators of the group $M(3,1)_{\mathrm{int}%
}$, the operators of spin projections $\hat{S}_{\mu \nu }$ are not
generators of the group $M(3,1)_{\mathrm{ext}}$, see (\ref{gen.R}) and 
(\ref{gen.L}), and therefore operators (\ref{10}) are functions of right (but not left)
generators of the Poincar\'{e} group.

For $SL(2,C)_{\mathrm{int}}$ we use two sets of commuting operators
corresponding to the reduction schemes 
\begin{eqnarray*}
&&SL(2,C)_{\mathrm{int}}\supset U(1)\times U(1), \\
&&SL(2,C)_{\mathrm{int}}\supset SU(2)\supset U(1).
\end{eqnarray*}%
In the first case two generators $\hat S^3_R$ and $\hat B^3_R$ of
the maximal commutative (Cartan) subgroup of $SL(2,C)_{\mathrm{int}}$
correspond to additive quantum numbers. In the second case the set includes $%
\hat S^3_R$ and Casimir operator $\hat{\mathbf{S}}_{R}^{2}$ of $SU(2)_{%
\mathrm{int}}$ with eigenvalues $S_R (S_R +1)$.

Therefore, we will consider two sets of 10 commuting operators on the group 
$M(3,1)$: 
\begin{eqnarray}
&&\hat{{W}}^{2},\;\hat{p}_{\mu },\;\hat{\mathbf{p}}\hat{\mathbf{S}}\;
(\hat{S}^{3}\,\,\hbox{in the rest frame}),\;\hat{\mathbf{S}}^{2}-\hat{\mathbf{B}}^{2},\;
\hat{\mathbf{S}}\hat{\mathbf{B}},\;\;\hat S^3_R,\;\hat B^3_R,  
\label{11} \\
&&\hat{W}^{2},\;\hat{p}_{\mu },\;\hat{\mathbf{p}}\hat{\mathbf{S}}\;
(\hat S^3\,\,\hbox{in the rest frame}),\;\hat{\mathbf{S}}^{2}-\hat{\mathbf{B}}^{2},\;
\hat{\mathbf{S}}\hat{\mathbf{B}},\;\;\hat{\mathbf{S}}_{R}^{2},\;\hat S^3_R,  
\label{31set-S}
\end{eqnarray}
including the Lubanski--Pauli operator $\hat{W}^{2}$,
four left generators $\hat{p}_{\mu }$ (the eigenvalue of the Casimir
operator $\hat{p}^2$, is evidently expressed through
their eigenvalues), and helicity $\hat{\mathbf{p}}\hat{\mathbf{S}}$,
expressed through the left generators. The Casimir operators 
$\hat{\mathbf{S}}^{2}-\hat{\mathbf{B}}^{2}$ and $\hat{\mathbf{S}}\hat{\mathbf{B}}$ 
of the subgroup $SL(2,C)_{\mathrm{int}}$ determine characteristics $j_{1},j_{2}$ of
the irrep $T_{[{j_{1}j_{2}}]}$ of the Lorentz group (see Appendix).

\subsection{Case of non-zero mass \label{S2-1}}

As is known, particles with non-zero mass are characterized by intrinsic
parity $\eta $ -- eigenvalue of the space reflection operator $P$ (which
defines the outer automorphism of the Poincar\'{e} group).

The action of space reflection $P$ on orientation variables is given by the
formula 
\begin{equation}
P:Z\rightarrow (Z^{\dagger })^{-1},\ \text{\textrm{or}}\ \left( 
\begin{array}{cc}
z^{1} & {\underline{z}}^{1} \\ 
z^{2} & {\underline{z}}^{2}%
\end{array}%
\right) \rightarrow \left( 
\begin{array}{cc}
-{\cc\uz}_{\dot{1}} & \cc z_{\dot{1}} \\ 
-{\cc\uz}_{\dot{2}} & \cc z_{\dot{2}}%
\end{array}%
\right) .  \label{13}
\end{equation}

The operator of the space reflection $\hat{P}$ anticommutes with the
operators $\hat{\mathbf{S}}\hat{\mathbf{B}}$ and $\hat B^3_R$.
Therefore, eigenfunctions of the latter three operators change their sign
under the action of $\hat{P}$. The set (\ref{31set-S}) is more convenient to
describe states of definite internal parity, because only one operator $\hat{%
\mathbf{S}}\hat{\mathbf{B}}$ from this set doesn't commute with $\hat{P}$.
Eigenvalues of $\hat{\mathbf{S}}\hat{\mathbf{B}}$ are proportional to $%
(j_{1}-j_{2})(j_{1}+j_{2}+1)$, see (\ref{SL2Ccas}). Thus, one can use
quantum numbers, corresponding to the set (\ref{31set-S}), to characterize
eigenstates of $\hat{P}$, with only one change (replacement the sign of 
$j_{1}-j_{2}$ by intrinsic parity $\eta $).

Taking into account the above said, for the massive fields on the Poincar\'{e} 
group, we will consider the following set of quantum numbers: 
\begin{equation}  \label{14}
p_{\mu },\;m^{2}=p_{\mu }p^{\mu }\neq 0,\;s,\;\eta ,\;
 {\mathbf{p}}\hat{\mathbf{S}}\;(S^3\; \hbox{in the rest frame}),\;
j_{1}+j_{2}, \; |j_{1}-j_{2}|,\; S_R,\; S^3_R.
\end{equation}
Along with the eigenvalues of the two Casimir operators, intrinsic parity 
$\eta$ labels irreps of the improper Poincar\'e group.

\subsection{Case of zero mass \label{S2-2}}

The massless case corresponds to the zero eigenvalue of the Casimir
operator, $\hat{{p}}^2 f(x,z) = 0$. The small group is a
three-parameter group of motions of a two-dimensional Euclidean space $E(2)$. 
This is a non-compact group, its unitary irreps are either
infinite-dimensional (the so-called representations with continuous spin) or
one-dimensional. In the latter case for representations of zero mass
(classified by irreps of $SO(2)\subset E(2)$) we have \cite{BogLoO90,EllDa79} 
\begin{equation}  \label{m0}
\hat W_\mu f(x,z) = \lambda \hat p_\mu f(x,z).
\end{equation}
In this case, irreps of the Poincar\'{e} group are labelled by the chirality 
$\lambda = j_{1}-j_{2}$ \cite{BarRa77}. Taking into account the identity 
$\hat {{W}}\hat {{p}}\equiv 0$, we obtain for square of Lubanski-Pauli operator 
\begin{equation*}
\hat {{W}}^2f(x,z)= (\hat {{W}}^2 + \lambda^2\hat {{p}}^2)f(x,z) = (\hat {{W}} - 
\lambda \hat {{p}})^2f(x,z)=0.
\end{equation*}

Let's write out the first equation of the system (\ref{m0}) for the
component $\hat W_0=\frac12 \varepsilon_{0\nu\rho\sigma}\hat p^\nu\hat
S^{\rho\sigma}= \hat{\mathbf p}\hat{\mathbf S}$: 
\begin{equation}  \label{m00}
\hat{\mathbf p}\hat{\mathbf S} f(x,z)=(j_1-j_2)\hat p_0 f(x,z).
\end{equation}
This equation establishes the equality of helicity and chirality, which significantly
limits the number of possible states of massless particles. 

For spatial components $\hat{W}^{k}=\hat{p}^{0}\hat{S}^{k}-\varepsilon ^{kli}\hat{p}^{i}\hat{B}^{l}$ we get 
\begin{equation}
(\hat{p}^{0}\hat{S}^{k}-\varepsilon ^{kji}\hat{p}^{i}\hat{B}^{j})f(x,z)=(j_{1}-j_{2})\hat{p}^{k}f(x,z).  
\label{m01}
\end{equation}%
In the case of representations $T_{[j\,0]}$ and $T_{[0\,j]}$, the equations 
(\ref{m01}) can be simplified by excluding boost operators $\hat B^k$.
Consider the operators $\hat M^k =\frac 12(\hat S^k-i\hat B^k)$, $\barM^k =-\frac 12(\hat S^k+i\hat B^k)$, see (\ref{MNgen}).
We have $\hat M^k f(x,z) =\frac 12(\hat S^k-i\hat B^k)=0$ for $T_{[0\,j]}$ and 
$\barM^k f(x,z) =-\frac 12(\hat S^k+i\hat B^k)=0$ for $T_{[j\,0]}$, see (\ref{MNgen}),
hence $\hat B^k= i \sign\!\lambda\ \hat S^k$,
\begin{equation}
\label{m02}
\left(\hat p^0\hat S^k - i \sign(j_1-j_2) \varepsilon_{kji}\hat p^i\hat S^j + (j_1-j_2)\hat p^k\right) f(x,z) =0.
\end{equation}
Using (\ref{m01}), one can exclude $\hat p^0$,
\begin{equation}
\label{m03}
\left((j_1-j_2)^{-1}\hat p^i\hat S^i\hat S^k - i \sign(j_1-j_2) \varepsilon_{kji}\hat p^i\hat S^j + (j_1-j_2)\hat p^k\right) f(x,z) =0.
\end{equation}
If the spin for a massive particle marks irreps of the Poincar\'{e} group
and is determined through the eigenvalue of the Casimir operator 
$\hat{{W}}^{2} $, then for massless particles irreps are marked by the
chirality, and the spin in this case is often defined as the modulus of
chirality. However, spin 1 particles can have chirality 1
or 0 in the massless limit; in this sense the identification of spin with
the modulus of chirality is not always correct.

For massless particles we arrive at a set of operators 
\begin{equation}
\hat{{p}}^{2}f=0,\;\hat{W}=\lambda \hat{p}\;(\hat{\mathbf{p}}\hat{%
\mathbf{S}}f=\lambda \hat p_0 f),\;\hat{p}_{\mu },\;\hat{\mathbf{S}}^{2}-%
\hat{\mathbf{B}}^{2},\;\hat{\mathbf{S}}\hat{\mathbf{B}},\; \hat B^3_R,\;\hat S^3_R.  \label{16}
\end{equation}%
The eigenvalues of the operators $\hat{{p}}^{2}$ and $\hat{{W}}^{2}$ are 
fixed and equal to 0, the helicity (function of the
left generators) is equal to the chirality (function of the right
generators) and labels the irreps of the Poincar\'{e} group.

Thus, in contrast to the massive case, instead of $10$ we have only $7$
different quantum numbers -- helicity (equal to the chirality $j_{1}-j_{2}$%
), three functions of the left generators (momentum components), three
functions of the right generators ($j_{1}+j_{2}$ and two internal quantum
numbers $B^3_R,\;  S^3_R$ or $S_R ,\;  S^3_R$ ).

In contrast to the massive case, described by the Dirac and Duffin-Kemmer
equations, associated with operators that do not commute with some
generators of $M(3,1)_{int}$ (right generators), and, therefore, associated
with broken inner symmetries, the massless case does not a priory imply such
a violation.

The case $m^{2}=0$ includes also functions on the group $f(z)$ that depend
only on the orientation variables and do not depend on the spatial variables 
$x^{\mu }$. They are classified according to the irreps of the Lorentz group
(the maximal set in this case includes $6$ operators -- two Casimir
operators $\hat{\mathbf{S}}^{2}-\hat{\mathbf{B}}^{2}$ and $\hat{\mathbf{S}}%
\hat{\mathbf{B}}$, the left operators $\hat{\mathbf{S}}^{2}$and $\hat S^3$%
, the right operators $\hat{\mathbf{S}}_{R}^{2}$ and $\hat S^3_R$).
However, these states cannot be localized in any region of the space, and,
therefore, they cannot be interpreted as particle states, so we will not
consider them here.

The case $m^{2}<0$ corresponds to tachyons, the small group $SU(1,1)$ is
non-compact, its unitary irreps are either one-dimensional or
infinite-dimensional. As in the case of zero mass, spin-finite-dimensional
representations are classified according to the irreps of the group $%
SO(2)\subset SU(1,1)$.

\section{$S^3_R$ charge\label{S3}}

The physical meaning of the left generators included in the maximal set of
operators is  well known; two of the four functions of the right generators
define the irrep of the Lorentz group $SL(2,C)_{int}$. Let's try to
understand the physical meaning of the remaining two operators $\hat{\mathbf{%
S}}_{R}^{2}$ and $\hat S^3_R$.

One can attempt to associate the charge corresponding to the right generator 
$\hat S^3_R$ of the Poincar\'{e} group to observable characteristics of
physical particles. Notice that although the internal quantum numbers
corresponding to the right generators do not change under the left
transformations, the discrete transformations (automorphisms of the Poincar%
\'{e} group) act on both right and left generators (see \cite{GitSh09} for
details). The known behavior under the discrete transformations may help one
to identify right characteristics with properties of physical particles.

In this regard, let's list the properties of the operator $\hat S^3_R$:

1. Its quantum number is additive.

2. It has integer eigenvalues for particles of integer spin and half-integer
eigenvalues for particles of half-integer spin.

3. It does not change the sign under the space reflection.

4. It changes the sign under the charge conjugation.

Then it follows from item 4 that this number equals to zero for real neutral
particle (i.e., particle that coincides its own antiparticle); at the same
time, real neutral particles with definite $S^3_R$ must to have an
integer spin.

If one considers $S^3_R$-charge of particles described by
finite-dimensional representations of the Lorentz group $T_{[j_{1},j_{2}]}$, 
$j_{1}+j_{2}=s$, then for particles of spin one-half two values are
possible: $1/2$ and $-1/2$; for particles of spin one three values are
possible: $1,0$ and $-1$. In particular, for a photon and $Z^{0}$-boson, as
for real neutral particles, we have $S^3_R=0$.

Let us see which eigenvalues $S^3_R$ can be associated with particles
described by eigenfunctions $f(x,z)$ of the operator $\hat S^3_R\ $,%
\begin{equation*}
\hat S^3_Rf(x,z)=S^3_Rf(x,z)\ .
\end{equation*}%
Since the sign of $\hat S^3_R$ changes under the charge conjugation,
particles and antiparticles must have $S^3_R$-charges of opposite signs.
For definiteness, let an electron $e^{-}$ has the $S^3_R$-charge $-\frac{%
1}{2}$; then a positron $e^{+}$ has $S^3_R=\frac{1}{2}$. For $W^{-}$, as
a charged particle, it is natural to expect $S^3_R=\pm 1$. Next, since $%
\tilde{\nu}_{e}$ admits only the values $\pm \frac{1}{2}$, the reaction $%
W^{-}\rightarrow e^{-}+\tilde{\nu}_{e}$ implies $S^3_R=-\frac{1}{2}$ for 
$\tilde{\nu}_{e}$ and $S^3_R=-1$ for $W^{-}$. Therefore, we have 
\begin{equation}
\begin{array}{lclcc}
S^3_R=\frac{1}{2}: & \nu _{e}\ ,\ e^{+}; & S^3_R=1: & W^{+}; &  \\ 
S^3_R=-\frac{1}{2}: & e^{-},\ \tilde{\nu}_{e}; & S^3_R=-1: & W^{-} & 
S^3_R=0:\gamma ,Z^{0},H; \\ 
&  &  &  & 
\end{array}
\label{s3rp1}
\end{equation}

Applying similar reasons to other families of fundamental fermions, we find
the following classification with respect to the sign of the quantum number $%
S^3_R$ :%
\begin{equation}
\begin{array}{lccccccc}
S^3_R=\frac{1}{2}: & \nu _{e} & \nu _{\mu } & \nu _{\tau } & \; & u & c
& t \\ 
S^3_R=-\frac{1}{2}: & e^{-} & \mu & \tau &  & d & s & b%
\end{array}
\label{s3rp2}
\end{equation}%
Therefore, the $S^3_R$-charge, whose sign changes under the charge
conjugation ${C}$, distinguishes not only particles and antiparticles but
also the \textquotedblleft up-down\textquotedblright\ components in doublets
of elementary fermions. This charge is conserved in any interactions, since
the carriers of electromagnetic and strong interactions are characterized by 
$S^3_R=0$, whereas we have already considered weak interaction case.

As a consequence of Eqs. (\ref{s3rp1}) and (\ref{s3rp2}), we may write the
following relation between $S^3_R$ and some other charges: 
\begin{equation}
S^3_R=\frac{L-B}{2}+Q\ ,  \label{s3rLBQ}
\end{equation}%
where $L,B$, and $Q$ are the lepton, baryon and electric charges,
respectively. Equation (\ref{s3rLBQ}) relates the \textquotedblleft
right\textquotedblright\ charge $S^3_R$ with observable characteristics
of the particles.

For the above-mentioned fundamental particles of spin one, the charge $%
S^3_R$ coincides with the electric charge. Taking into account that $%
(B-L)/2=\bar{Q}$, where $\bar{Q}$ -- average electric charge of
corresponding lepton $(\nu ,e^{-})$ or quark $(d,u)$ doublet one can rewrite
Eq. (\ref{s3rLBQ}) as $S^3_R=Q-\bar{Q}$.

Formula (\ref{s3rLBQ}) is closely related to the concept of $R$-parity. In
1961 Michel and Lur\c{c}at \cite{LurMi61} observed that for all the known
particles with integer $B$ there holds the relation $(-1)^{B+L}=(-1)^{2S}$,
in other words, the number $B+L+2S$ is always even. Later, this observation
resulted in the concept of $R$-parity, being positive for all the known
particles, and defined as 
\begin{equation}  \label{Rpar}
R=(-1)^{3(B-L)+2S},
\end{equation}
where the multiplier 3 is introduced to include quarks in the consideration.

We have already noted that the spectra of right and left spin operators are
not independent, in particular (see item 2 above), 
$S^3_R$-charge and spin $S$ must both be either integers or half-integers,
\begin{equation}
\label{SS3}
(-1)^{2S^3_R }=(-1)^{2S}.
\end{equation}
Due to (\ref{s3rLBQ}) we have $(-1)^{L-B+2Q}=(-1)^{2S}$, therefore, the
value 
\begin{equation}
\label{R1}
R_1=(-1)^{2S^3_R +2S}=(-1)^{L-B+2Q+2S}
\end{equation}
is always positive. For integer $Q$ we have $R_1=R$, for fractional $Q$
(quarks), it is not difficult to verify that $(-1)^{L-B+2Q}=(-1)^{3(B-L)}$.
We see, that for all known particles $R_1=R$.

Thus, consideration of the field on the Poincar\'e group together with the
empirical formula (\ref{s3rLBQ}) leads to the connection of spin with
internal quantum numbers, expressed in the concept of $R$-parity.

We also note that $S^3_R$ coincides with the projection of the weak
isospin $T_{3}$ for left components of particles and right of antiparticles.
In fact, in the massless limit, the formula (\ref{s3rLBQ}) turns into the
well-known relation $T_{3}=-Y/2+Q$, where $Y$ is a weak hypercharge. The
possible connection of $SU(2)_{int}$ with the weak gauge group is
considered in Ref. \cite{GitSh11}.

In the next section we describe functions $f(x,z)$, polynomial in $z$.
Considering polynomials of fixed degree $n$ in $z$, we will actually
consider the irreps of the Lorentz group $T_{[j_{1}\;j_{2}]}$ with $%
j_{1}+j_{2}=n/2$. According to the reduction formula (\ref{SL2Cred}), the
values $j_{1}$ and $j_{2}$ define a possible range of spin values $S$, 
$S_R$ and projections $S^{3}$ and $S^3_R$. 
\begin{equation}
\begin{array}{|c|c|c|}
\hline
\text{\textrm{Irreps of} }SL(2,C)_{int} & S,S_R & \text{\textrm{projections} 
}S^{3},\;S^3_R \\ \hline
T_{[1/2\;0]},\;T_{[0\;1/2]} & 1/2 & \pm 1/2 \\ \hline
T_{[1\;0]},\;T_{[0\;1]} & 1 & 0,\pm 1 \\ \hline
T_{[1/2\;1/2]} & 0,\;1 & 0,\pm 1 \\ \hline
T_{[3/2\;0]},\;T_{[0\;3/2]} & 3/2 & \pm 1/2;\pm 3/2 \\ \hline
T_{[1\;1/2]},\;T_{[1/2\;1]} & 1/2,\;3/2 & \pm 1/2;\pm 3/2 \\ \hline
\end{array}
\label{20}
\end{equation}

\section{Massless case\label{S4}}

\subsection{$\boldsymbol{j_{1}+j_{2}=0}$, chirality $0$}

The case of ordinary scalar functions $f(x)=e^{-ipx^0+ipx^3}$, which do not
depend on orientation variables $z$. Corresponds to a scalar particle.

\subsection{$\boldsymbol{j_{1}+j_{2}=1/2}$, chirality $\boldsymbol{\pm 1/2}$}

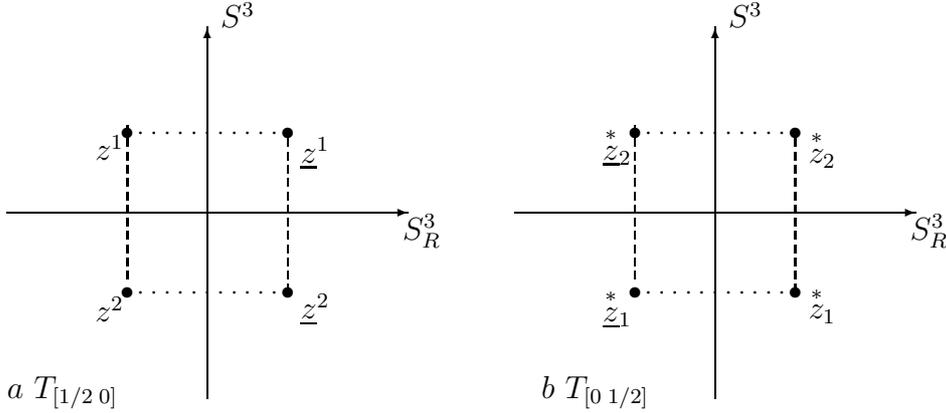
\begin{figure}[th]
\caption{The weight diagrams of the representations $T_{[1/2\;0]}$ and 
$T_{[0\;1/2]}$ of $SL(2,C)_{\mathrm{ext}}\times SL(2,C)_{\mathrm{int}}$.
Vertical lines join states related by transformations of $SL(2,C)_{\mathrm{%
ext}}$, horizontal lines join states related by transformations of $SL(2,C)_{%
\mathrm{int}}$. }
\label{rr2} 
\begin{picture}(380,150)
\newsavebox{\spinhalfa}
\savebox{\spinhalfa}(200,100)[lb]
{
\put(0,70){\vector (1,0){150}}
\put(75,0){\vector (0,1){140}}
\put(105,40){\circle*{4}}
\put(45,100){\circle*{4}}
\put(45,40){\circle*{4}}
\put(105,100){\circle*{4}}
\put(148,60){$S^3_R $}
\put(80,140){$S^3$}
\multiput(45,100)(5,0){12}{\circle*{1}} 
\multiput(45,40)(5,0){12}{\circle*{1}}
\multiput(45,100)(0,-5){12}{\line(0,1){3}}
\multiput(105,40)(0,5){12}{\line(0,1){3}}
}
\put(10,0){\usebox{\spinhalfa}}
\put(10,0){$a$ $T_{[1/2\; 0]}$}
\put(120,30){$\uz^2$}
\put(43,90){$z^1$}
\put(43,29){$z^2$}
\put(120,89){$\uz^1$}
\put(200,0){\usebox{\spinhalfa}}
\put(210,0){$b$ $T_{[0 \; 1/2]}$}
\put(310,30){$\cc z_1$}
\put(233,90){$\cc \uz_2$}
\put(233,29){$\cc \uz_1$}
\put(310,89){$\cc z_2$}
\end{picture}
\end{figure}

Let us consider functions linear in $z$. Irreps enter into the decomposition
of the left (right) regular representation with a multiplicity equal to the
dimension of the representation; see \cite{BarRa77}. This can be clearly
illustrated on the Fig. \ref{rr2}. It shows weight diagrams of the
representations $T_{[1/2\;0]}$ (chirality $\frac{1}{2}$) and $T_{[0\;1/2]}$
(chirality $-\frac{1}{2}$) of the dimension 2. When acting from the left,
the spinors $(z_{1},z_{2})$ and $(\uz_{1},\uz_{2})$ are transformed identically 
and correspond to equivalent representations; when acting from the right they
are transformed identically $(z_{1},\uz_{1})$ and $(z_{2},\uz_{2})$. If we 
consider simultaneously the action on the left and the right (external and internal
transformations ), then all four states are mixed.

In total we have 8 states (orientation variables 
$z^{\alpha },\uz^{\alpha },\cc z^{\alpha },\cc\uz^{\alpha }$), corresponding
to four different particles (in other words, states that cannot be
translated into each other by transformations of the laboratory reference
system). If states inside quadruplets are connected by left or right
transformations of the Poincar\'{e} group, then quadruplets are connected to
each other by discrete transformations (space reflection, complex (charge)
conjugation), which are external automorphisms of the group.
Particle-antiparticle pairs correspond to complex conjugate pairs $z^{\alpha
},\cc z^{\alpha }$ and ${\underline{z}}{\vphantom{z}}^{\alpha },\cc\uz%
^{\alpha }$; inside the pairs the signs of the chirality $j_{1}-j_{2}$ and
the charge $S_{R}^{3}$ are opposite.

\begin{figure}[ht]
\caption{The weight diagram of the representation $T_{[1/2\; 0]}\oplus
T_{[0\; 1/2]}$ of $SL(2,C)_{\mathrm{int}}$. The dotted line joins states
related by transformations of $SL(2,C)_{\mathrm{int}}$. a) States with
definite chirality, functions (\protect\ref{LRm}). b) States with definite
parity $\protect\eta$, functions (\protect\ref{etam}).}
\label{rr24}
\savebox{\spinhalfa}(160,180)[lb] {\ \put(0,70){\vector (1,0){150}}
\put(75,0){\vector (0,1){140}} 
\put(105,40){\circle*{4}} \put(108,30){$\uz-\cc z,\,e^+$} \put(45,100){%
\circle*{4}} \put(0,90){$z-\cc \uz,\;e^-$} \put(45,40){\circle*{4}}
\put(0,29){$z+\cc \uz,\;\tilde \nu$} \put(105,100){\circle*{4}} \put(108,89){%
$\uz+\cc z,\;\nu$} \put(148,60){$S^3_R $} \put(80,140){$\eta$} \put(0,0){$b$}
\multiput(45,40)(5,0){12}{\circle*{1}} \multiput(45,100)(5,0){12}{\circle*{1}%
} }
\begin{picture}(380,150)
\put(0,70){\vector (1,0){150}}
\put(75,0){\vector (0,1){140}}
\put(105,40){\circle*{4}}
\put(108,30){$\cc z,\,\nu_L$}
\put(45,100){\circle*{4}}
\put(18,90){$\cc \uz,\,e^-_L$}
\put(45,40){\circle*{4}}
\put(18,29){$z,\,\tilde\nu_R$}
\put(105,100){\circle*{4}}
\put(108,89){$\uz,\,e^+_R$}
\put(148,60){$S^3_R $}
\put(80,140){$-iB^3_R $}
\put(0,0){$a$}
\put(200,0){\usebox{\spinhalfa}}
\multiput(45,100)(5,-5){12}{\circle*{1}} 
\multiput(45,40)(5,5){12}{\circle*{1}}
\end{picture}
\end{figure}
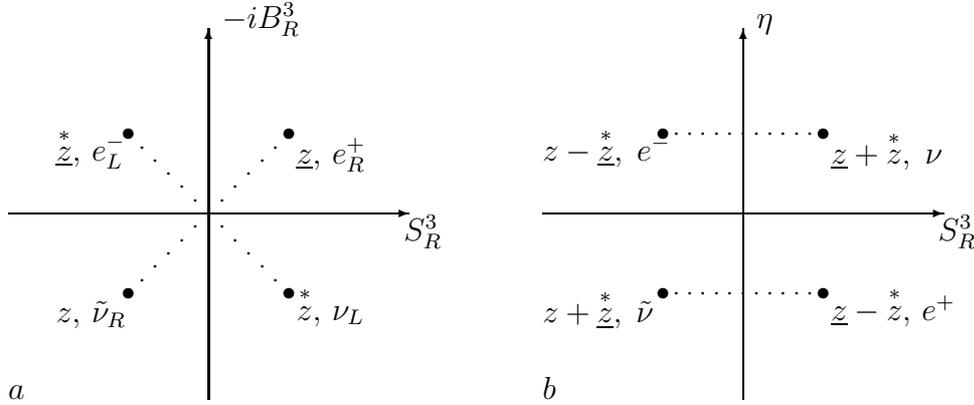

The decomposition of the representation of the Poincar\'{e} group in the
space of linear in $z$ functions $f(x,z)$ is carried out using a set of
operators (\ref{16}). Consider the equation 
$\hat{\mathbf p}\hat{\mathbf S}f(x,z)=(j_{1}-j_{2})\hat{p}_{0}f(x,z)$, establishing the equality of the
chirality and the helicity, from the set. For a massless particle with a
certain momentum $p^\mu=(p^0,0,0,p^3)$, $p^3=\pm p^0$, we have 
$p^{3}S^{3}=(j_{1}-j_{2})p^{0}$; for a positive $p^{0}$ signs of $p^{3}$ and 
$S^{3}$ must coincide for particles, and be opposite for antiparticles
(positive and negative helicity, right-handed particles and left-handed
antiparticles). Thus, the classification of functions on the Poincar\'{e}
group gives two left-handed particles and two right-handed antiparticles,
each having one chiral state, 
\begin{equation}
\begin{array}{lcc}
        &   S^3_R =-1/2 &  S^3_R =1/2 \\
R\;(j_1-j_2=1/2): \quad & e^{-ipx^0+ipx^3}z^1 \quad   & e^{-ipx^0+ipx^3}\uz^1  \\
L\;(j_1-j_2=-1/2):\quad & e^{-ipx^0-ipx^3}\cc\uz_{\dot 1} \quad & e^{-ipx^0-ipx^3}\cc z_{\dot 1}   \\
\end{array}
\label{LRm}
\end{equation}%
These two particle/antiparticle pairs differ in signs of the charge 
$S_{R}^{3}$, states with the same chirality are connected by a group 
transformation $SU(2)_{int}$, see Fig. \ref{rr24}a. 

We see that in the space of the functions on the Poincar\'{e} group,
particles of spin one half appear not in pairs, but in quadruplets.

\subsection{$\boldsymbol{j_{1}+j_{2}=1}$, chirality $\boldsymbol{\pm 1}$}

Let us consider functions on the group that are quadratic in $z$.

From the variables $z,{\underline{z}}{\vphantom{z}}$ as well as from 
$\cc z,\cc\uz$, 10
different quadratic combinations can be constructed. 9 of them correspond to
the irreps $T_{[1\,0]}$ and $T_{[0\,1]}$ (chirality $j_{1}-j_{2}=\pm 1$),
two combinations are invariants, since, due to unimodularity, $\det Z=z^{1}{%
\underline{z}}{\vphantom{z}}^{2}-z^{2}{\underline{z}}{\vphantom{z}}^{1}=1$
is a Lorentz scalar.

The representations $T_{[1\,0]}$ and $T_{[0\,1]}$ correspond to figures
similar to the one Fig. \ref{rr2}, but each of them has not 4, but 9 states.

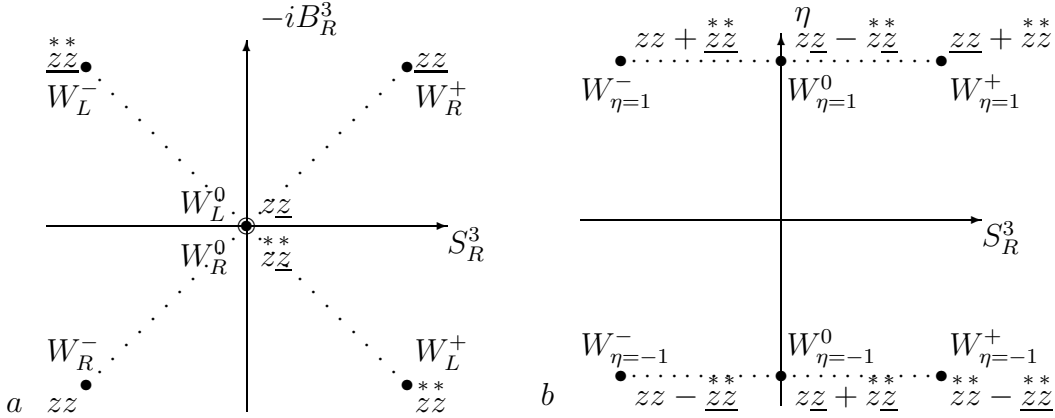
\begin{figure}[ht]
\caption{The weight diagrams of the representation $T_{[1\; 0]}\oplus
T_{[0\; 1]}$ of $SL(2,C)_{\mathrm{int}}$, $j_1-j_2=\pm 1$ (a), $\protect\eta=\pm 1$
(b). The dotted lines join states related by the transformations $%
SL(2,C)_{int}$.}
\label{hir1}\newsavebox{\spinonea} \savebox{\spinonea}(160,165)[lb] {\
\put(0,70){\vector (1,0){150}} \put(75,0){\vector (0,1){140}} 
\put(20,135){$zz+\cc \uz \cc \uz$} \put(0,115){$W^-_{\eta=1}$} \put(138,0){$%
\cc z \cc z-\cc\uz\cc\uz$} \put(15,11){\circle*{4}} \put(135,11){\circle*{4}%
} \put(15,130){\circle*{4}} \put(135,130){\circle*{4}} \put(20,0){$%
zz-\cc\uz\cc\uz$} \put(0,19){$W^-_{\eta=-1}$} \put(138,135){$\uz\uz+\cc z\cc
z$} \put(138,19){$W^+_{\eta=-1}$} \put(138,115){$W^+_{\eta=1}$} \put(75,11){%
\circle*{4}} \put(75,130){\circle*{4}} \put(80,135){$z\uz-\cc z\cc \uz$}
\put(80,0){$z\uz+\cc z\cc \uz$} \put(77,19){$W^0_{\eta=-1}$} \put(77,115){$%
W^0_{\eta=1}$} \put(150,60){$S^3_R $} \put(80,145){$\eta$} \put(-15,0){$b$}
\multiput(15,130)(5,0){25}{\circle*{1}} \multiput(15,11)(5,0){25}{\circle*{1}%
} }
\par
\begin{picture}(360,165)
\put(0,70){\vector (1,0){150}}
\put(75,0){\vector (0,1){140}}
\put(0,130){$\cc \uz \cc \uz$}
\put(0,115){$W^{-}_L$}
\put(138,0){$\cc z \cc z$}
\put(15,10){\circle*{4}}
\put(135,10){\circle*{4}}
\put(15,130){\circle*{4}}
\put(135,130){\circle*{4}}
\put(0,0){$zz$}
\put(0,19){$W^{-}_R$}
\put(138,130){$\uz\uz$}
\put(138,19){$W^{+}_L$}
\put(138,115){$W^{+}_R$}
\put(75,70){\circle*{4}}
\put(75,70){\circle{6}}
\put(80,75){$z\uz$}
\put(80,55){$\cc z\cc \uz$}
\put(50,75){$W^0_L$}
\put(50,55){$W^0_R$}
\put(150,60){$S^3_R $}
\put(80,145){$-iB^3_R $}
\put(-15,0){$a$}
\multiput(15,130)(5,-5){25}{\circle*{1}}
\multiput(15,10)(5,5){25}{\circle*{1}}

\put(200,0){\usebox{\spinonea}}
\end{picture}
\end{figure}

By restoring the laboratory indices, one can easily observe that each point
of the weight diagram (Fig.\ref{hir1}) corresponds to three states
(according to the number of possible spin projections), that transform under
3-dimensional irreps of $SL(2,C)_{ext}$. In particular, for states with $%
S^3_R=0$ among four pairwise products 3 functions correspond to the
chirality one, namely $z^{1}{\underline{z}}{\vphantom{z}}^{1},\;z^{2}{%
\underline{z}}{\vphantom{z}}^{2}$, and$\;z^{1}{\underline{z}}{\vphantom{z}}%
^{2}+z^{2}{\underline{z}}{\vphantom{z}}^{1}$.

Making a reduction to the compact group $SU(2)_{int}$, we also obtain two
triplets: left and right, corresponding to the diagonals on Fig.\ref{hir1}a,
or triplets with a fixed parity, Fig.\ref{hir1}b.

For a massless particle with a certain momentum $p^\mu=(p^0,0,0,p^3)$, $p^3=\pm p^0$, we have 
$p^{3}S^{3}=(j_{1}-j_{2})p^{0}$; for a positive $p^{0}$ signs of $p^{3}$ and 
$S^{3}$ must coincide for particles, and be opposite for antiparticles
(positive and negative helicity, right-handed particles and left-handed
antiparticles). 
Thus, we have 6 states (particles) $W_{L}^{\pm }$, $W_{L}^{0}$, $W_{R}^{\pm
} $, $W_{R}^{0}$, corresponding to three values of $S^3_R$ and two
values of the chirality; for the states with $p^\mu=(p,0,0,p)$ we have 
\begin{equation}  \label{LRm1}
\begin{array}{lccc}
        &   S^3_R =-1 &  S^3_R =0 &  S^3_R =1 \\
R\;(j_1-j_2=1): \quad & e^{-ipx^0+ipx^3}z^1z^1 \quad  & e^{-ipx^0+ipx^3}z^1\uz^1 \quad  & e^{-ipx^0+ipx^3}\uz^1\uz^1  \\
L\;(j_1-j_2=-1):\quad & e^{-ipx^0-ipx^3}\cc\uz_{\dot 1}\cc\uz_{\dot 1} \quad &  e^{-ipx^0-ipx^3}\cc z_{\dot 1}\cc\uz_{\dot 1} \quad  
& e^{-ipx^0-ipx^3}\cc z_{\dot 1}\cc z_{\dot 1}\\
\end{array}
\end{equation}
By direct substitution, it is not difficult to make sure that the states (\ref{LRm1}) satisfies not only (\ref{m00}), but also (\ref{m02}).

In this case, particles with the same $S^3_R$ differ only by signs of
the chirality/helicity and can be considered as different polarization
states of the same particle.

\subsection{$\boldsymbol{j_{1}+j_{2}=1}$, chirality $0$}

Let us now consider the remaining quadratic combinations of $z$, which are
products of $z,{\underline{z}}{\vphantom{z}}$ to the complex conjugates 
$\cc z,\cc \uz$. In total, there are 16 such combinations and they belong 
to the $T_{[\frac{1}{2}\frac{1}{2}]}$ representation of the Lorentz group.
For this representation, all the states are characterized by the chirality $%
j_{1}-j_{2}=0$. During a reduction to the compact group $SU(2)_{int}$ (or $%
SU(2)_{ext}$) we obtain a triplet and a singlet.

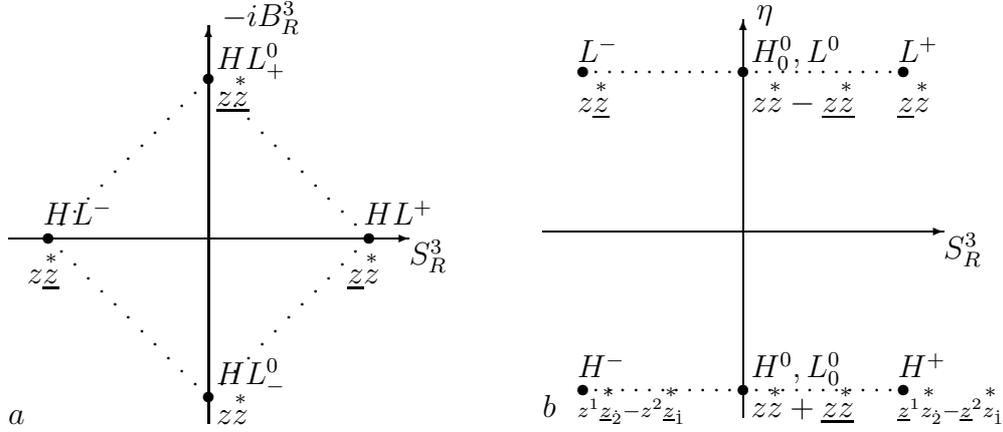
\begin{figure}[ht]
\caption{The weight diagrams of the representation $T_{[\frac 12\;\frac 12]}$
of $SL(2,C)_{\mathrm{int}}$, chirality $j_1-j_2=0$. On Fig.a the dotted line joins
states related by transformations of $SL(2,C)_{int}$, on Fig.b the
dotted line joins states related by transformations of $SU(2)\in
SL(2,C)_{int}$.}
\label{rrhir0}
\newsavebox{\spinoneb} 
\savebox{\spinoneb}(160,160)[lb] {
\put(0,70){\vector (1,0){150}} 
\put(75,0){\vector (0,1){150}} 
\put(13,115){$z \cc \uz$} 
\put(133,115){$\uz\cc z$} 
\put(78,0){$z \cc z+\uz \cc \uz$} 
\put(78,115){$z \cc z-\uz \cc \uz$} 
\put(75,10){\circle*{4}}
\put(15,10){\circle*{4}} 
\put(135,10){\circle*{4}} 
\put(75,130){\circle*{4}}
\put(15,130){\circle*{4}} 
\put(135,130){\circle*{4}} 
\put(150,60){$S^3_R $}
\put(80,150){$\eta$} 
\put(0,0){$b$} 
\multiput(15,130)(5,0){25}{\circle*{1}}
\multiput(15,10)(5,0){25}{\circle*{1}} 
\put(13,134){$L^-$} \put(78,134){$H^0_0,L^0$} \put(133,134){$L^+$} 
\put(13,15){$H^-$} \put(78,15){$H^0,L^0_0$} \put(133,15){$H^+$} 
\put(13,0){$\scriptstyle{z^1\cc\uz_{\dot2}-z^2\cc\uz_{\dot1}}$} 
\put(133,0){$\scriptstyle{\uz^1\cc z_{\dot2}-\uz^2\cc z_{\dot1}}$} 
} 
\begin{picture}(360,170)
\put(0,70){\vector (1,0){150}}
\put(75,0){\vector (0,1){150}}
\put(7,53){$z \cc \uz$}
\put(127,53){$\uz\cc z$}
\put(78,0){$z \cc z$}
\put(78,120){$\uz \cc \uz$}
\put(75,10){\circle*{4}}
\put(75,130){\circle*{4}}
\put(15,70){\circle*{4}}
\put(135,70){\circle*{4}}
\put(150,60){$S^3_R $}
\put(80,150){$-iB^3_R $}
\put(0,0){$a$}
\multiput(15,70)(5,-5){12}{\circle*{1}}
\multiput(15,70)(5,5){12}{\circle*{1}}
\multiput(135,70)(-5,-5){12}{\circle*{1}}
\multiput(135,70)(-5,5){12}{\circle*{1}}
\put(13,75){$HL^-$}
\put(78,134){$HL^0_+$}
\put(133,75){$HL^+$}
\put(78,15){$HL^0_-$}
\put(200,0){\usebox{\spinoneb}}
\end{picture}
\end{figure}

Each point on the Fig. \ref{rrhir0}a corresponds to 4 states, namely 3
vector components and a scalar with respect to the rotation group $%
SU(2)_{ext}$. States with a fixed charge $S^3_R$ can be distinguished
either by $B^3_R$, or by $S_R $ and the parity $\eta$. 

For the states with a certain momentum $p^\mu=(p^0,0,0,p^3)$, $p^3=\pm p^0$,
the equation (\ref{m00}) gives $\hat p^3\hat S^3 f(x,z)=0$.
Thus, the only one components remains with $S^{3}=0$, 
\begin{equation*}  
f(x,z)=e^{-ipx^0\pm ipx^3}\varphi(z), \qquad  \hat S^3 \varphi(z)=0. 
\end{equation*}  
These states must also satisfy the equations (\ref{m01}) for spatial components $W_i$,
$W^i f(x,z)=p^0\hat S^i - \varepsilon_{ijk}p^j\hat B^k f(x,z) =0$. 
The third equation is fulfilled identically, the first two equations give $p^3=-p^0$ for functions of the form $z^1\cc z^1$ and
$p^3=p^0$ for functions of the form $z^2\cc z^2$. Since these functions are connected by rotation from
$SU(2)_{ext}$, then we have the following 4 states corresponding to 4 particles,
\begin{equation}  
\label{chir0}
\begin{array}{ccc}
 S^3_R =-1 &  S^3_R =0 &  S^3_R =1 \\
 e^{-ipx^0 - ipx^3}z^1\cc \uz^{\dot1} & 
 e^{-ipx^0 - ipx^3}\uz^1\cc \uz^{\dot1} &
 e^{-ipx^0 - ipx^3}\uz^1\cc z^{\dot1}\\
 & e^{-ipx^0 - ipx^3}z^1\cc z^{\dot1} & \\
\end{array}
\end{equation}  
The two states with $S^3_R =0$ differ in the values of $B^3_R $.

Let us consider scalars with respect to the group $SU(2)_{ext}$: 
\begin{eqnarray}
&&H^-:\; z^1\cc\uz_{\dot1}+z^2\cc\uz^{\dot2}, \quad
  H^+:\; \uz^1\cc z_{\dot1}+\uz^2\cc z^{\dot2}, \quad \\
&&	
  H^0_-:\;  z^1\cc z_{\dot1}+z^2\cc z^{\dot 2}, \quad  
	H^0_+:\; \uz^1\cc \uz_{\dot1}+\uz^2\cc \uz^{\dot2}, \quad 
\end{eqnarray}%
-- eigenfunctions of spin projection operators $\hat{S}_{k}$ with zero
eigenvalues. Their linear combinations are the triplet $H^{-},H^{0},H^{+}$ ($%
S_R =1$, $\eta =1$) and the singlet $H_{0}^{0}$ ($S_R =0$, $\eta =-1$)
with respect to $SU(2)_{int}$, 
\begin{equation}
H^0:\; z^1\cc z^{\dot1}+z^2\cc z^{\dot 2} - \uz^1\cc \uz^{\dot1}-\uz^2\cc \uz^{\dot2}, \qquad 
H^0_0:\; z^1\cc z^{\dot1}+z^2\cc z^{\dot 2} + \uz^1\cc \uz^{\dot1}+\uz^2\cc \uz^{\dot2},
\end{equation}
see Fig.\ref{rrhir0}b. All this functions  satisfy the equation (\ref{m00})
identically, for them both spin projections and chirality are equal to zero.
However, the set of these states is not invariant with respect to the boost operators
contained in the equations (\ref{m01}),
which convert these states into vectors with respect to $SU(2)_{ext}$.

The results of the consideration of massless case can be presented in the form of a Table 
\ref{massless}.

\begin{table}[!ht]
\caption{Classification of scalar functions and its interpretation, massless
case.}
\label{massless}%
\begin{equation*}
\begin{array}{|c|c|c|r|c|l|}
\hline
\text{chirality} &  & \multicolumn{2}{|c|}{\text{irreps of}} & \text{%
dimension of } &  \\ 
j_1-j_2 & j_1\;j_2 & \multicolumn{2}{|c|}{SU(2)_{int}\times SU(2)_{ext}} & 
SU(2)_{int}\text{ irreps} & \mathrm{particles} \\ 
&  & \multicolumn{2}{|c|}{\text{and its dimension}} &  &  \\ \hline
\frac12 & \frac12\; 0 & \qquad T_\frac12\times T_\frac12 \qquad & 4 & 2 & 
e^+_R, \bar\nu_R \\ \hline
-\frac12\;\; & 0\; \frac12 & T_\frac12\times T_\frac12 & 4 & 2 & e^-_L, \nu_L
\\ \hline
1 & 1\; 0 & T_1\times T_1 & 9 & 3 & W^+_R,W^-_R,W^0_R \\ \hline
-1\;\; & 0\; 1 & T_1\times T_1 & 9 & 3 & W^+_L,W^-_L,W^0_L \\ \hline
0 & \frac12\; \frac12 &  & 16 &  &  \\ 
&  & (T_1+T_0)\times T_1 & 12 & 3 & HL^+,HL^0,HL^- \\ 
&  & (T_1+T_0)\times T_0 & 4  & 1 & HL^{0}_0 \\ 
\hline
\end{array}%
\end{equation*}%
\end{table}

For fixed $j_{1}$ and $j_{2}$, the number of particles (states not connected
by transformations $SL(2,C)_{ext}$) is given by the dimension of the irrep
of the $SU(2)_{int}$ group. Each of the listed massless particles can be in
one chiral state, for which the projection of the spin onto the direction of
motion coincides with $j_{1}-j_{2}$.

For the case $j_{1}+j_{2}>1$ the consideration scheme is similar. For
example, let us consider polynomials of degree 3 in $z$, namely, $%
j_{1}+j_{2}=3/2$, and chirality is $\pm 1/2$ and $\pm 3/2$.

\subsection{$\boldsymbol{j_{1}+j_{2}=3/2}$, chirality $\pm 3/2$}

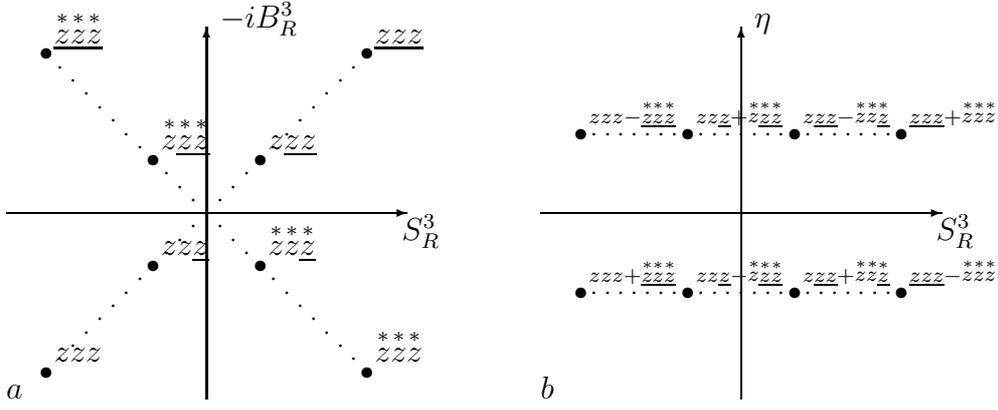
\begin{figure}[ht]
\newsavebox{\spintd} \savebox{\spintd}(160,180)[lb] {\ \put(0,70){\vector
(1,0){150}} \put(75,0){\vector (0,1){140}} 
\put(135,100){\circle*{4}} \put(95,100){\circle*{4}} \put(55,100){\circle*{4}%
} \put(15,100){\circle*{4}} \put(135,40){\circle*{4}} \put(95,40){\circle*{4}%
} \put(55,40){\circle*{4}} \put(15,40){\circle*{4}} 
\put(138,104){$\scriptstyle{\uz\uz\uz + \cck z\cck z\cck z}$} \put(98,104){$%
\scriptstyle{z\uz\uz - \cck z\cck z\cck \uz}$} \put(58,104){$\scriptstyle{%
zz\uz + \cck z\cck \uz\cck \uz}$} \put(18,104){$\scriptstyle{zzz -
\cck\uz\cck\uz\cck\uz}$} \put(138,44){$\scriptstyle{\uz\uz\uz - \cck z\cck
z\cck z}$} \put(98,44){$\scriptstyle{z\uz\uz + \cck z\cck z\cck \uz}$}
\put(58,44){$\scriptstyle{zz\uz - \cck z\cck \uz\cck \uz}$} \put(18,44){$%
\scriptstyle{zzz + \cck\uz\cck\uz\cck\uz}$} \put(148,60){$S^3_R $}
\put(80,140){$\eta$} \put(0,0){$b$} \multiput(15,40)(5,0){25}{\circle*{1}}
\multiput(15,100)(5,0){25}{\circle*{1}}} 
\begin{picture}(380,150)
\put(0,70){\vector (1,0){150}}
\put(75,0){\vector (0,1){140}}
\put(135,130){\circle*{4}}
\put(95,90){\circle*{4}}
\put(55,50){\circle*{4}}
\put(15,10){\circle*{4}}

\put(135,10){\circle*{4}}
\put(95,50){\circle*{4}}
\put(55,90){\circle*{4}}
\put(15,130){\circle*{4}}

\put(138,14){$\cc z\cc z\cc z$}
\put(98,54){$\cc z\cc z\cc \uz$}
\put(58,94){$\cc z\cc \uz\cc \uz$}
\put(18,134){$\cc\uz\cc\uz\cc\uz$}

\put(18,14){$zzz$}
\put(58,54){$zz\uz$}
\put(98,94){$z\uz\uz$}
\put(138,134){$\uz\uz\uz$}

\put(148,60){$S^3_R $}
\put(80,140){$-iB^3_R $}

\put(0,0){$a$}
\put(200,0){\usebox{\spintd}}
\multiput(15,130)(5,-5){25}{\circle*{1}} 
\multiput(15,10)(5,5){25}{\circle*{1}}
\end{picture}
\caption{The weight diagram of the representation $T_{[3/2\; 0]}\oplus
T_{[0\; 3/2]}$ of $SL(2,C)_{\mathrm{int}}$. The dotted line joins states
related by transformations of $SL(2,C)_{\mathrm{int}}$. a) States with
definite chirality $j_1-j_2=\pm 3/2$. b) States with definite parity $%
\protect\eta$.}
\label{rr320}
\end{figure}

The representations $T_{[\frac{3}{2}0]}$ and $T_{[0\frac{3}{2}]}$ of the
Lorentz group have the dimension 4 and are associated with 16 monomials of
the third degree in $z,{\underline{z}}{\vphantom{z}}$. In total, there are 8
states (particles) not connected by transformations $SL(2,C)_{int}$,
differing by the values $S^3_R$ $(-3/2,-1/2,1/2,3/2)$ and by the
chirality sign $(-3/2,3/2)$, corresponding to 8 points on the Fig. \ref%
{rr320}a. Each point of the weight diagram (Fig. \ref{rr320}) corresponds to
four states (according to the number of possible spin projections).

The equation (\ref{m00}) for a fixed momentum directed along the $x^{3}$
axis has the form $p_{3}\hat{S}^{3}f(x,z)=(j_{1}-j_{2})p_{0}f(x,z)$;
accordingly, for a fixed chirality and the charge $S^3_R$, one component
remains, characterized by $S^{3}=j_{1}-j_{2}$. These states can be
interpreted as 8 particles or as 4 particles in 2 polarization states.

\subsection{$\boldsymbol{j_{1}+j_{2}=3/2}$, chirality $\pm 1/2$}

The representations $T_{[1\frac{1}{2}]}$ and $T_{[\frac{1}{2}1]}$ of the
Lorentz group have the dimension 6 and each of them corresponds to 36
monomials of third degree. Being reduced to a compact subgroup $SU(2)$, they
split into 4-dimensional $T_{3/2}$ and 2-dimensional $T_{1/2}$
representations. These states can be interpreted as 24 particles or 12
particles in two polarization states.

\begin{figure}[ht]
\caption{The weight diagram of the representation $T_{[1\; 1/2]}\oplus
T_{[1/2\; 1]}$ of $SL(2,C)_{\mathrm{int}}$. The dotted line joins states
related by transformations of $SL(2,C)_{\mathrm{int}}$. a) States with
definite chirality $j_1-j_2=\pm 1/2$. b) States with definite parity $%
\protect\eta$.}
\label{rr321}
\savebox{\spintd}(160,180)[lb] {\ \put(0,70){\vector (1,0){150}}
\put(75,0){\vector (0,1){140}} 
\put(135,100){\circle*{4}} \put(95,100){\circle*{4}} \put(55,100){\circle*{4}%
} \put(15,100){\circle*{4}} \put(135,40){\circle*{4}} \put(95,40){\circle*{4}%
} \put(55,40){\circle*{4}} \put(15,40){\circle*{4}} \put(138,104){$%
\scriptstyle{\uz\cck z(\uz + \cck z)}$} \put(98,104){$\scriptstyle{\uz\cck
z(\cck\uz - z)}$} \put(58,104){$\scriptstyle{z\cck\uz (\uz + \cck z)}$}
\put(18,104){$\scriptstyle{z\cck\uz (z - \cck\uz)}$} \put(98,30){$%
\scriptstyle{z\cck z\cck z + \uz \uz\cck \uz}$} \put(58,90){$\scriptstyle{%
zz\cck z + \uz\cck \uz\cck \uz}$} \put(138,44){$\scriptstyle{\uz\cck z(\uz -
\cck z)}$} \put(98,44){$\scriptstyle{\uz\cck z(\cck\uz + z)}$} \put(58,44){$%
\scriptstyle{z\cck\uz (\uz - \cck z)}$} \put(18,44){$\scriptstyle{z\cck\uz
(z + \cck\uz)}$} \put(98,90){$\scriptstyle{z\cck z\cck z - \uz \uz\cck \uz}$%
} \put(58,30){$\scriptstyle{zz\cck z - \uz\cck \uz\cck \uz}$} \put(148,60){$%
S^3_R $} \put(80,140){$\eta$} \put(0,0){$b$} \multiput(15,40)(5,0){25}{%
\circle*{1}} \multiput(15,100)(5,0){25}{\circle*{1}} } 
\begin{picture}(380,160)
\put(0,70){\vector (1,0){160}}
\put(75,0){\vector (0,1){150}}
\put(135,90){\circle*{4}}

\put(95,130){\circle*{4}}
\put(55,130){\circle*{4}}
\put(95,90){\circle*{4}}
\put(55,50){\circle*{4}}

\put(15,50){\circle*{4}}

\put(135,50){\circle*{4}}

\put(95,10){\circle*{4}}
\put(55,10){\circle*{4}}
\put(95,50){\circle*{4}}
\put(55,90){\circle*{4}}

\put(15,90){\circle*{4}}

\put(58,134){$\uz\cc\uz\cc\uz$}
\put(98,134){$\uz\uz\cc\uz$}
\put(138,94){$\uz\uz\cc z$}
\put(18,94){$z\cc\uz\cc\uz$}

\put(98,94){$\uz\cc \uz\cc z$}
\put(58,94){$z \uz\cc\uz$}
\put(98,54){$z \uz\cc z$}
\put(58,54){$z \cc z\cc\uz$}

\put(138,54){$\uz\cc z\cc z$}
\put(18,54){$z z\cc\uz$}
\put(58,14){$z z\cc z$}
\put(98,14){$z \cc z\cc z$}

\put(158,60){$S^3_R $}
\put(80,147){$-iB^3_R $}

\put(0,0){$a$}
\put(200,0){\usebox{\spintd}}
\multiput(15,50)(5,5){17}{\circle*{1}}
\multiput(55,10)(5,5){17}{\circle*{1}}
\multiput(15,90)(5,5){8}{\circle*{1}}
\multiput(95,10)(5,5){8}{\circle*{1}}
\multiput(55,50)(5,5){8}{\circle*{1}}

\multiput(95,10)(-5,5){17}{\circle*{1}} 
\multiput(135,50)(-5,5){17}{\circle*{1}}
\multiput(55,90)(5,-5){8}{\circle*{1}}
\multiput(55,10)(-5,5){8}{\circle*{1}}
\multiput(135,90)(-5,5){8}{\circle*{1}}

\end{picture}
\end{figure}
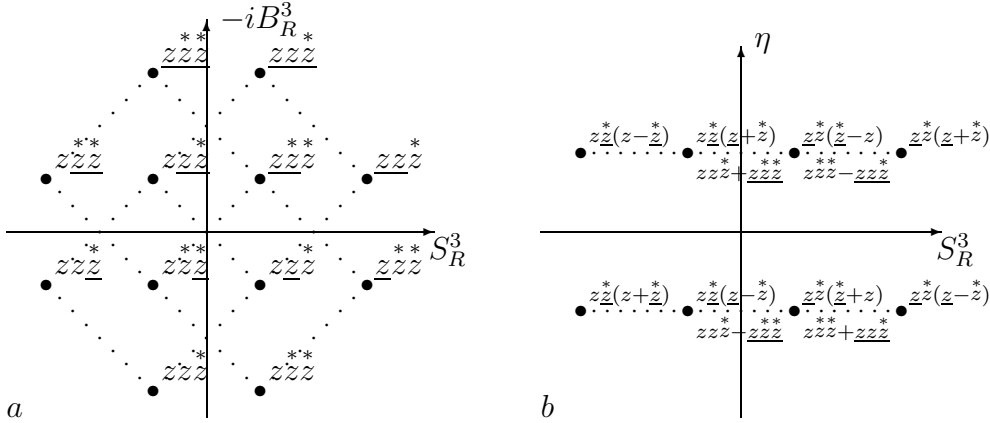

\section{Relativistic wave equations for massless particles\label{S5}}

The operators $\hat W_\mu$ and $\hat S_k$ in equations (\ref{m0})--(\ref{m02}%
), acting in the space of functions $f(x,z)$ on the group, are differential
in $z$ and for all representations have the same form. By writing $f(x,z)$
as polynomials of $z$ with coefficients $\psi_{n}(x)$ and comparing the
coefficients at powers of $z$, we can write equations (\ref{m0})--(\ref{m02}%
) as matrix equations for $\psi_{n}(x)$, and their form will be different in
different representations.

1. $j_{1}+j_{2}=1/2$, the chirality $\pm 1/2$, $T_{[\frac{1}{2}0]}$ and $%
T_{[0\frac{1}{2}]}$. Let us write down the functions corresponding to $%
S^3_R=1/2$: 
\begin{eqnarray*}
&&f(x,\uz)=\psi_1(x)\uz^1 + \psi_2(x)\uz^2, \quad \psi_R(x)=(\psi_1(x)\; \psi_2(x))^T,
\\
&&f(x,\cc z)=\psi^{\dot 1}(x)\cc z_{\dot 1} + \psi^{\dot 2}(x)\cc z_{\dot 2}, \quad \psi_L(x)=(\psi^{\dot 1}(x)\; \psi^{\dot 2}(x))^T
\end{eqnarray*}%
Substituting these forms into Eq. (\ref{m00}) and comparing coefficients  at 
${\uz}_{i}$ and at $\cc z_{\dot{i}}$, we obtain a pair of Weyl equations,
\begin{equation}  
\label{weyl}
(\hat{p}_{0}-\boldsymbol{\sigma}\hat{\mathbf p})\psi _{R}=0,\quad 
(\hat{p}_{0}+\boldsymbol{\sigma}\hat{\mathbf p})\psi _{L}=0.
\end{equation}
The same equations we obtain for the case $S^3_R=-1/2$. For chiralities 
$\pm 1/2$ the equations (\ref{m03}) for spatial components of $\hat W_\mu$ will be fulfilled automatically if 
(\ref{m00}) is fulfilled. To see this, it is enough to note that in the case under consideration 
$\mathbf S=\frac12\boldsymbol{\sigma}$.

2. $j_{1}+j_{2}=1$, the chirality $\pm 1$, $T_{[10]}$ and $T_{[01]}$. The
representations $T_{[10]}$ and $T_{[01]}$ of the Lorentz group are
three-dimensional complex conjugate representations by the matrices $SO(3,C)$%
. For the complex functions $\psi _{k}=E_{k}+iH_{k}$ transforming under $%
T_{[10]}$, the combination $\sum \psi _{k}^{2}=\sum (E_{k}+iH_{k})^{2}=%
\mathbf{E}^{2}-\mathbf{H}^{2}+2i\mathbf{EH}$ is conserved. Let us write
functions $f(x,z,{\underline z}{\vphantom{z}})$ and $f(x,\cc z,\cc\uz)$,
corresponding to $S^3_R=0$ and connected by space reflection: 
\begin{eqnarray*}
&f(x,z,\uz)=\psi_{1}(x)i(z^2\uz^2-z^1\uz^1) + \psi_{2}(x)(-z^1\uz^1-z^2\uz^2) + \psi_{3}(x)i(z^1\uz^2+z^2\uz^1), \quad \\
&\xi=\psi_R(x)=(\psi_{1}\, \psi_{2}\, \psi_{3})^T; \\
&f(x,\cc z,\cc\uz)=\psi^{1}(x)i(\cc z_{\dot 2}\cc \uz_{\dot 2}-\cc z_{\dot 1}\cc \uz_{\dot 1}) + \psi^{2}(x)(-\cc z_{\dot 1}\cc \uz_{\dot 1}-\cc z_{\dot 2}\cc \uz_{\dot 2}) + \psi^{3}(x)i(\cc z_{\dot 1}\cc \uz_{\dot 2}+\cc z_{\dot 2}\cc \uz_{\dot 1}), \quad \\
&\eta=\psi_L(x)=(\psi^{1}\, \psi^{2}\, \psi^{3})^T.
\end{eqnarray*}

Substituting these forms into Eqs. (\ref{m00}) and (\ref{m03}) and comparing
expressions at the orientation variables, we obtain Maxwell equations in the
Majorana form: 
\begin{eqnarray}
i{\frac {\partial \xi }{\partial t}}={\hat{\mathbf{p}}\mathbf{S}}\xi, \quad {%
\hat{\mathbf{p}}}{\boldsymbol{\xi }}=0,\quad i{\frac {\partial \eta }{%
\partial t}}=-{\hat{\mathbf{p}}}\mathbf{S}\eta , \quad {\hat{\mathbf{p}}}{%
\boldsymbol{\eta }}=0,\quad  \label{22a}
\end{eqnarray}%
where $(S^i)_{kl}=\varepsilon_{ikl}$, 
\begin{equation*}
S^{1}={%
\begin{pmatrix}
0 & 0 & 0 \\ 
0 & 0 & -i \\ 
0 & i & 0%
\end{pmatrix}%
},\ S^{2}={%
\begin{pmatrix}
0 & 0 & i \\ 
0 & 0 & 0 \\ 
-i & 0 & 0%
\end{pmatrix}%
},\ S^{3}={%
\begin{pmatrix}
0 & -i & 0 \\ 
i & 0 & 0 \\ 
0 & 0 & 0%
\end{pmatrix}%
}.
\end{equation*}%
The column ($\mathbf{\xi ~\eta }$)$^{T}$ is considered as the photon wave function \cite{BerLiP,Bialy94}.

3. $j_{1}+j_{2}=1$, the chirality $0$, $T_{[\frac{1}{2}\frac{1}{2}]}$.
Let us write down the functions corresponding to $S^3_R=1/2$: 
\begin{equation*}
\begin{array}{l}
f(x,z)=\psi_\mu(x)\sigma^\mu_{\;\;\alpha\dot\beta} \uz^\alpha \cc z^{\dot\beta}
\\
=
\psi_0(x)(\uz^1 \cc z^{\dot1}+\uz^2 \cc z^{\dot2}) +
\psi_1(x)(\uz^1 \cc z^{\dot2}+\uz^2 \cc z^{\dot1}) +
\psi_2(x)i(\uz^2 \cc z^{\dot1}-\uz^1 \cc z^{\dot2}) +
\psi_3(x)(\uz^1 \cc z^{\dot1}-\uz^2 \cc z^{\dot2}).  			
\end{array}
\end{equation*}
Substituting these forms into Eq. (\ref{m00}) and comparing
coefficients at the orientation variables, we obtain
\begin{equation} 
\label{22b}
\hat {\mathbf p}{{\mathbf S} }{{\psi}}=0, \quad \psi(x)=(\psi_1(x),\psi_2(x),\psi_3(x))^T,
\end{equation}
where $(S^i)_{kl}=\varepsilon_{ikl}$, or $\nabla\times \psi=0$ in vector notations.
Functions $\psi_0(x)$, corresponding to scalar $H^{+}$, are not included in the equation, because $\hat S_k (\cc z^1\uz^{\dot1}+z^2\cc\uz^{\dot2})=0$.
Equation (\ref{m01}) can be written in matrix form,
\begin{equation} 
\label{22c}
({\hat p^0}\tilde S^i-\varepsilon_{ijk}\hat p^j B^k) \Psi (x)  =0, \quad \Psi(x)=(\psi_0(x),\psi_(x))^T,
\end{equation}
where matrices $B^k$ have two nonzero elements $(B^k)_{0k}=(B^k)_{k0}=i$, 
and $\tilde S^i$ are obtained from $S^i$ by adding the first zero row and column. 
The latter equation is equivalent to the system of equation (\ref{22b}) and three equations
$\frac{\partial\psi_k}{\partial t}-\frac{\partial\psi_0}{\partial x^k}=0$.

\section{Massive case\label{S6}}

Here we consider eigenfunctions of the maximal set of 10 commuting operators
(12). Since the representations $T_{[j_{1}j_{2}]}$ are $T_{[j_{2}j_{1}]}$
are connected by a space reflection, then, as was mentioned above, to
construct states with a certain internal parity instead of chirality $%
j_{1}-j_{2}$ one has to use the modulus $|j_{1}-j_{2}|$ and the internal
parity given by the set of numbers (\ref{14}): the mass $m$, the spin $S$,
the parity $\eta $, the momentum $\mathbf{p},$ the spin projection $S_{3}$,
parameters of irrep of the Lorentz group $j_{1}+j_{2}$, $|j_{1}-j_{2}|$, $%
S_R $ and its projection $S^3_R$.

We will consider the rest frame $(p_{0}=m,\;p_{k}=0)$; in this frame, the
operator $\hat{{W}}^{2}$ reads: 
\begin{equation}
\hat{{W}}^{2}f(x,z)=-m^{2}S(S+1)f(x,z),  \label{22}
\end{equation}
where $S(S+1)$ is an eigenvalue of operator $\hat{\mathbf S}^2$.

The irreps $T_{[j0]}$ and $T_{[0j]}$ of the group $SL(2,C)$ when reduced to
the compact subgroup $SU(2)$ remain irreducible $T_{j}$ of the dimension $%
2j+1$. For such representations $S=j$. For other irreps of $SL(2,C)$, upon
the reduction, the representation becomes reducible and, for a given $%
j_{1},j_{2}$, we obtain different values of the spin and irreps of the
Poincar\'{e} group.

\subsection{$\boldsymbol{j_{1}+j_{2}=0}$, spin 0}

The case of ordinary scalar functions $f(x)=e^{-imx^0}$, which do not depend
on orientation variables $z$. Corresponds to a scalar particle with $\eta=1$.

\subsection{$\boldsymbol{j_1+j_2=1/2}$, spin 1/2}

Consider the representations $T_{[\frac{1}{2}0]}$ and $T_{[0\frac{1}{2}]}$.
The spin $S=1/2$, in the rest frame 
\begin{equation}
\begin{array}{lcc}
        &   S^3_R =-1/2 &  S^3_R =1/2 \\
\eta=1 \quad  & e^{-imx^0}(z^\alpha - \cc \uz_\da) \quad & e^{-imx^0}(\uz^\alpha + \cc z_\da) \\
\eta=-1\quad  & e^{-imx^0}(z^\alpha + \cc \uz_\da) \quad & e^{-imx^0}(\uz^\alpha - \cc z_\da) \\
\end{array}
\label{etam}
\end{equation}

Each of 4 functions corresponds to a point on the Fig. \ref{rr24}b.
Different signs of the spin projection $S^{3}=\pm 1/2$ correspond to two
possible values of the index $\alpha $ in Eq. (\ref{etam}).

There are 8 states, 4 particles; the particle states in the rest frame
differ in two spin projection.

\subsection{$\boldsymbol{j_1+j_2=1}$, spin 1}

The representations $T_{[10]}$ and $T_{[01]}$ are connected by space
reflection, $S=j_{1}+j_{2}=1$. Each point in Fig. \ref{hir1}b corresponds to
3 states with different spin projections. 
\begin{equation}  \label{etam1}
\begin{array}{lccc}
& S^3_R =-1 & S^3_R =0 & S^3_R =1 \\ 
\eta=\pm 1 & e^{-imx^0}(z^\alpha z^\beta \pm\cc \uz_{\dot\alpha} \cc \uz%
_{\dot\beta}) & e^{-imx^0}(z^{(\alpha}{\underline z}{\vphantom{z}}^{\beta)}
\mp\cc z_{({\dot\beta}}\cc \uz_{{\dot\alpha})}) & e^{-imx^0}({\underline z}{%
\vphantom{z}}_\alpha{\underline z}{\vphantom{z}}_\beta \pm\cc z_{\dot\alpha}%
\cc z_{\dot\beta}) 
\end{array}%
\end{equation}
Here, the parentheses of the indexes indicate symmetrization.

There are 18 states, 3 particles; the particle states differ in parity $\eta
=\pm 1$ and 3 spin projection values in the rest frame.

\subsection{$\boldsymbol{j_1+j_2=1}$, spin 0 and 1}

The representation $T_{[\frac{1}{2}\frac{1}{2}]}$, 16 states in total. The
four-dimensional representation, when reduced to a compact subgroup,
decomposes into a scalar $(S=0)$ and a vector $(S=1)$.

For $S=0$, we have 4 particles, a singlet $H_{0}^{0}$ and a triplet $%
H^{-},H^{0},H^{+}$ with respect to $SU(2)_{int}$. In the rest frame 
\begin{equation}  \label{etamH}
\begin{array}{lccc}
             &   S^3_R =-1 &  S^3_R =0 &  S^3_R =1 \\
S^R=1 \quad  & e^{-imx^0}(z^1\cc\uz^{\dot1}+z^2\cc\uz^{\dot2}) \quad & e^{-imx^0}(z^1\cc z^{\dot1}+z^2\cc z^{\dot 2} - \uz^1\cc \uz^{\dot1}-\uz^2\cc \uz^{\dot2}) &
 e^{-imx^0}(\uz^1\cc z^{\dot1}+\uz^2\cc z^{\dot2})\\
S^R=0 \quad  &  \quad & e^{-imx^0}(z^1\cc z^{\dot1}+z^2\cc z^{\dot 2} + \uz^1\cc \uz^{\dot1}+\uz^2\cc \uz^{\dot2}) & \\
\end{array}
\end{equation}

Scalars represent 4-plet of the $SL(2,C)_{int}$. Being reduced to a compact
subgroup $SU(2)_{int}$, they split into the triplet $H^{-},H^{0},H^{+}$ and
the singlet $H_{0}^{0}$ (see Fig. \ref{rrhir0}b). In the triplet $%
S^3_R=-1,0,+1$, and, according to the empirical formula (\ref{s3rLBQ}),
the same values of electrical charge. With respect to the operator $\hat{M}%
_{k}$ algebra (see \ref{MNgen}) the scalars form two doublets $%
H^{-},H_{-}^{0}$ and $H_{+}^{0},H^{+}$ (see Fig. \ref{rrhir0}a).

We note, that quantum numbers of the $H^{0}_0$ state (zero electric charge
and the parity $\eta =1$) coincide with those of the Higgs boson.

\bigskip For $S=1$, we also have 4 particles, a singlet $L_{0}^{0}$ and a
triplet $L^{-},L^{0},L^{+}$ with respect to $SU(2)_{int}$. 
\begin{equation}  \label{etamL}
\begin{array}{lccc}
        &   S^3=-1 &  S^3=0 & S^3=1 \\
L^-,\; S^3_R =\!-1 &	
e^{-imx^0}z^1 \cc \uz^{\dot2} & 
e^{-imx^0}(z^1\cc\uz^{\dot1}-z^2\cc\uz^{\dot2}) & 
e^{-imx^0}z^2 \cc \uz^{\dot1} \\ 			
L^+,\; S^3_R =1\; &	
e^{-imx^0}\cc z^1 \uz^{\dot2} & 
e^{-imx^0}(\cc z^1\uz^{\dot1}-z^2\cc\uz^{\dot2}) & 
e^{-imx^0}\cc z^2 \uz^{\dot1} \\ 			
L^0,\; S^3_R =0\; &	
e^{-imx^0}(z^1 \cc z^{\dot2} - \uz^1 \cc \uz^{\dot2})& 
e^{-imx^0}(z^1\cc z^{\dot1}\!-\!z^2\cc z^{\dot 2}\! + \!\uz^1\cc \uz^{\dot1}\!-\!\uz^2\cc \uz^{\dot2}) & 
e^{-imx^0}(z^2 \cc z^{\dot1} - \uz^2 \cc \uz^{\dot1})\\ 			
L^0_0,\; S^3_R =0\; &	
e^{-imx^0}(z^1 \cc z^{\dot2} + \uz^1 \cc \uz^{\dot2})& 
e^{-imx^0}(z^1\cc z^{\dot1}\!-\!z^2\cc z^{\dot 2} \! - \!\uz^1\cc \uz^{\dot1}\!+\!\uz^2\cc \uz^{\dot2}) & 
e^{-imx^0}(z^2 \cc z^{\dot1} + \uz^2 \cc \uz^{\dot1})\\ 			
\end{array}
\end{equation}
The case $S^3_R=0$  (fields $L^{0}$, $L_{0}^{0}$) corresponds to a truly neutral hypothetical particle,
which is called the notoph in the literature \cite{OgiPo67,KalRa74,Chizh11} (in this case, the
corresponding potentials are tensors transforming according to the representations $T_{[10]}$ and $T_{[01]}$).

Thus, just 8 particles (states not connected by transformations 
$SL(2,C)_{ext}$) are associated with the representation 
$T_{\frac{1}{2}}\times T_{\frac{1}{2}}$.

The above can be summarized in the Table \ref{massiv}. 
Each of the listed particles can be in several spin states, their number
coincides with the dimension of the $SU(2)_{ext}$ representation.

\begin{table}[!ht]
\caption{Classification of functions and their interpretation, $j_1+j_2\le 1$%
, $m \neq 0$.}
\label{massiv}%
\begin{equation*}
\begin{array}{|c|c|c|c|c|r|c|l|}
\hline
\mathrm{Spin} & \eta & |j_1-j_2| & \mathrm{Irreps\; of} & 
\multicolumn{2}{|c|}{\text{Irreps of}} & \text{dimension} & \mathrm{%
particles} \\ 
&  &  & SL(2,C)_{int} & \multicolumn{2}{|c|}{SU(2)_{int}\times SU(2)_{ext}}
& \text{of } SU(2)_{int} &  \\ 
&  &  &  & \multicolumn{2}{|c|}{\text{and its dimension}} & \text{irreps} &  \\ \hline
\frac12 & 1 & \frac12 & T_{[\frac12\; 0]},T_{[0,\;\frac12]} & 
T_\frac12\times T_\frac12 & 4 & 2 & e^-, \nu \\ \hline
\frac12 & -1\;\; & \frac12 & T_{[\frac12\; 0]},T_{[0,\;\frac12]} & 
T_\frac12\times T_\frac12 & 4 & 2 & e^+, \bar\nu \\ \hline
1 & \pm 1\;\; & 1 & T_{[1\; 0]},T_{[0,\;1]} & T_1\times T_1,\; T_1\times T_1
& 18 & 3,3 & W^+,W^-,W^0 \\ \hline
&  & 0 & T_{[\frac12\; \frac12]} &  & 16 &  &  \\ 
1 & 1 & 0 &  & T_1\times T_1 & 9 & 3 & L^+,L^-,L^0 \\ 
1 & -1\;\; & 0 &  & T_1\times T_0 & 3 & 1 & L^{00} \\ 
0 & -1 & 0 &  & T_0\times T_1 & 3 & 3 & H^+,H^0,H^- \\ 
0 & 1\;\; & 0 &  & T_0\times T_0 & 1 & 1 & H^{00} \\ \hline
\end{array}%
\end{equation*}%
\end{table}

\subsection{$\boldsymbol{j_1+j_2=3/2}$, spin 1/2 and 3/2}

The representations $T_{[\frac{3}{2}\,0]}$ and $T_{[0\,\frac{3}{2}]}$ are
related by space reflection, $S=j_{1}+j_{2}=3/2$. There are 32 states in
total and 8 particles (states that are not connected by transformations $%
SL(2,C)_{int}$, differing by the values $S^3_R$ $(\pm 1/2,\pm 3/2)$ and
by the parity, corresponding to 8 points on the Fig. \ref{rr320}b. Each
point of the weight diagram (Fig. \ref{rr320}b) corresponds to four states
(according to the number of possible spin projections).

The representations $T_{[1\,\frac{1}{2}]}$ and $T_{[\frac{1}{2}\,1]}$ are
related by a space reflection, $S=j_{1}+j_{2}=1$. There are 72 states in
total. These are 12 spin $3/2$ particles, differing by the parity, by the
values of $S_R$ $(1/2,3/2)$ and $S^3_R$ $(\pm 1/2,\pm 3/2)$, and 12 spin 
$1/2$ particles, differing by the parity, by the values of $S_R$ $(1/2,3/2)$
and $S^3_R$ $(\pm 1/2,\pm 3/2).$

\begin{table}[!ht]
\caption{Classification of scalar functions, $j_1+j_2=3/2$, $m\ne 0$.}
\label{massiv32}%
\begin{equation*}
\begin{array}{|c|c|c|c|c|r|c|}
\hline
\mathrm{Spin} & \eta & |j_1-j_2| & \mathrm{Irreps\; of} & 
\multicolumn{2}{|c|}{\text{Irreps of}} & \text{dimension of } \\ 
&  &  & SL(2,C)_{int} & \multicolumn{2}{|c|}{SU(2)_{int}\times SU(2)_{ext}}
& SU(2)_{int}\text{ irreps} \\ 
&  &  &  & \multicolumn{2}{|c|}{\text{and its dimension}} &  \\ \hline
3/2 & \pm 1\;\; & \frac32 & T_{[\frac32\; 0]},T_{[0,\;\frac32]} & 
T_\frac32\times T_\frac32,\; T_\frac32\times T_\frac32 & 32 & 4,4 \\ \hline
&  & \frac 12 & T_{[\frac32\;\frac 12]},T_{[\frac 12\;\frac32]} &  & 72 & 
\\ 
3/2 & 1 &  &  & T_\frac32\times T_\frac32, T_\frac32\times T_\frac32 & 36 & 4,4 \\ 
3/2 & -1\;\; &  &  & T_\frac32\times T_\frac 12, T_\frac32\times T_\frac 12 & 16 & 2,2 \\ 
1/2 & -1\;\; &  &  & T_\frac 12\times T_\frac32 T_\frac 12\times T_\frac32 & 16 & 4,4 \\ 
1/2 & 1 &  &  & T_\frac 12\times T_\frac 12, T_\frac 12\times T_\frac 12 & 8 & 2,2 \\ \hline
\end{array}%
\end{equation*}%
\end{table}


\section{The ultra-relativistic limit\label{S7}}

Let us now consider functions corresponding to a massive particle, moving
along the axis $x^{3}$. They can be obtained from functions in the rest
frame, which are characterized by a certain internal parity with the help of
a Lorentz transformation $U\in SL(2,C)$, 
\begin{eqnarray}
&&P=UP_{0}U^{\dagger },\quad Z=UZ_{0},\quad \hbox{ where}\;\;P_{0}=%
\mathop{{\rm diag}}\{m,m\},\quad U=\mathop{{\rm diag}}\{e^{a},e^{-a}\}, 
\notag  \label{sol2a} \\
&&p_{0}=m\cosh 2a,\quad p_{3}=m\sinh 2a,\quad e^{\pm a}=\sqrt{(p_{0}\pm
p_{3})/m}.
\end{eqnarray}%
For the orietation variables we have: 
\begin{equation*}
(z^{1},{\underline{z}}{\vphantom{z}}^{1},\cc z_{2},\cc \uz_{2})\rightarrow
e^{a}(z^{1},{\underline{z}}{\vphantom{z}}^{1},\cc z_{2},\cc \uz_{2}),\quad
(z^{2},{\underline{z}}{\vphantom{z}}^{2},\cc z_{1},\cc \uz_{1})\rightarrow
e^{-a}(z^{2},{\underline{z}}{\vphantom{z}}^{2},\cc z_{1},\cc \uz_{1}).
\end{equation*}%
If we apply these transformations to functions (\ref{etam})--(\ref{etamL})
and consider the ultra-relativistic limit, then only states for which
helicity is equal to chirality will remain in it.

For example, applying these transformations to the state with $S^3_R =-1/2$, $%
\eta=1$ from (\ref{etam}), we find 
\begin{equation}  \label{sol2}
f(x,z)\to f^{\prime i(k_0x^0+k_3x^3)}\left[ C_1(z^1e^a-\cc\uz_{\dot
1}e^{-a}) + C_2(z^2e^{-a}-\cc\uz_{\dot 2}e^a)\right],
\end{equation}
where the first term in the square brackets corresponds to $S^3=1/2$, and
the second term corresponds to $S^3=-1/2$. In the ultra-relativistic case
with a positive $a$ (i.e. with $k_3>0$) there remain only two components, 
\begin{equation}
f^{\prime }(x,{z}) \approx e^{i(k_0x^0+k_3x^3)}\left( C_1 z^1e^a - C_2\cc\uz%
_{\dot 2}e^a\right).
\end{equation}
The first term corresponds to $S^3>0$ and chirality $\lambda=1/2$, the
second term corresponds to $S^3<0$ and chirality $\lambda=-1/2$. We see,
that in the ultra-relativistic limit with $p_0>0$ the signs of chirality and
helicity $\hat p\hat S$ are the same.

As should be expected, the results of this analysis of the
ultra-relativistic case coincide with the conclusions on the chirality and
helicity of a particle with spin 1/2, obtained on a basis of the Dirac
equation.

For states (\ref{etam1}) of spin 1 (representations of $T_{[10]}$ and $%
T_{[01]}$) in the ultra-relativistic limit also remain only 2 terms with the
same signs of helicity and chirality.

For states (\ref{etamL}) of spin 1 (representation $T_{[\frac12,\frac12]}$),
in the ultra-relativistic limit there remains 1 term with $S^3=0$ and zero
chirality.

\section{First order left-invariant equations\label{S8}}

If in the case of massless particles, first-order equations in $\hat{p}^{\mu
}$ appear within the framework of the classification of functions with
respect to the irreps of the Poincar\'{e} group, then for massive particles
we have two second-order eigenvalue equations for the Casimir operators $%
\hat{\boldsymbol{p}}^{2}$ and $\hat{\boldsymbol{W}}^{2}$. For massive
particles, it is impossible to construct first-order equations using only
the generators of the Poincar\'{e} group. The desired equations must be
invariant under the transformations $M(3,1)_{ext}$, and for particles with a
certain parity, also under the space reflection. In this case, the
invariance with respect to $M(3,1)_{int}$ is not required, the internal
symmetries can be broken.

From the orientation variables $z$ one can construct operators that are
transformed as vectors with respect to $M(3,1)_{ext}$ and $M(3,1)_{int}$,
preserving the degree of polynomials in $z$; see Ref. \cite{GitSh09}, 
\begin{eqnarray}
&&\hat{\Gamma}^{\mu {\underline{n}}}=\frac{1}{2}\left( \bar{\sigma}^{\mu {%
\dot{\alpha}}\beta }\sigma _{\;\;{\underline{b}}\dot{\underline{a}}}^{{%
\underline{n}}}\cc z_{{\dot{\alpha}}}^{\;\;\dot{\underline{a}}}\partial
_{\beta }^{\;\;{\underline{b}}}+\sigma _{\;\;\beta {\dot{\alpha}}}^{\mu }%
\bar{\sigma}^{{\underline{n}}\dot{\underline{a}}{\underline{b}}}z_{\;\;{%
\underline{b}}}^{\beta }\partial _{\;\;\dot{\underline{a}}}^{\dot{\alpha}%
}\right) ,  \label{24} \\
&&\hat{\underline{\Gamma }}{\vphantom{\Gamma}}^{\mu {\underline{n}}}=\frac{1%
}{2}\left( \bar{\sigma}^{\mu {\dot{\alpha}}\beta }\sigma _{\;\;{\underline{b}%
}\dot{\underline{a}}}^{{\underline{n}}}\cc z_{{\dot{\alpha}}}^{\;\;\dot{%
\underline{a}}}\partial _{\beta }^{\;\;{\underline{b}}}-\sigma _{\;\;\beta {%
\dot{\alpha}}}^{\mu }\bar{\sigma}^{{\underline{n}}\dot{\underline{a}}{%
\underline{b}}}z_{\;\;{\underline{b}}}^{\beta }\partial _{\;\;\dot{%
\underline{a}}}^{\dot{\alpha}}\right) ,  \label{25}
\end{eqnarray}%
with one external (left) and one internal (right) index. It is only the
operators $\hat{p}_{\mu }\hat{\Gamma}^{\mu \underline{0}}$ and $\hat{p}_{\mu
}\hat{\underline{\Gamma }}{\vphantom{\Gamma}}^{\mu \underline{k}}$, $%
\underline{k}=1,2,3$, that are invariant under the space reflection, and the
associated 4 equations possess solutions with a definite inner parity.

Let us consider the following equation:%
\begin{equation}
\hat{p}^{\mu }\hat{\Gamma}^{\mu \underline{0}}f(x,z)=\varkappa ^{\underline{0%
}}f(x,z),  \label{26}
\end{equation}%
where 
\begin{eqnarray}
&&2\hat\Gamma^{\mu \underline{0}}  =  \bar \sigma ^{\mu{\dot\alpha}\beta} (%
\cc z_{{\dot\alpha}}^{\;\;\dot{\underline{1}}}\partial_{\beta}^{\;\;%
\underline{1}} + \cc z_{{\dot\alpha}}^{\;\;\dot {\underline{2}}%
}\partial_{\beta}^{\;\;\underline{2}}) + \sigma^{\mu}_{\;\; \beta{\dot\alpha}%
} ( z^{\beta}_{\;\;\underline{1}}\partial^{\dot \alpha}_{\;\;\dot{\underline{%
1}}} + z^{\beta}_{\;\;\underline{2}}\partial^{\dot \alpha}_{\;\;\dot {%
\underline{2}}})  \notag \\
&& =  (\bar \sigma ^{\mu{\dot\alpha}\beta} \cc{\underline z}_{{\dot\alpha}%
}\partial/\partial z^{\beta} + \sigma^{\mu}_{\;\; \beta{\dot\alpha}}
z^{\beta}\partial/\partial \cc {\underline z}_{\dot \alpha}) + (\bar \sigma
^{\mu{\dot\alpha}\beta} \cc z_{{\dot\alpha}}\partial/\partial {\underline z}%
^{\beta} + \sigma^{\mu}_{\;\; \beta{\dot\alpha}} {\underline z}%
^{\beta}\partial/\partial \cc z_{\dot \alpha}).
\end{eqnarray}
The set of the operators $\hat{S}^{\mu \nu }$ and $\hat{\Gamma}^{\mu 
\underline{0}}$ satisfies the commutation relations of the group $SO(3,2)$.

Obviously, the equation (\ref{26}) is not invariant under the right boosts,
since the operator $\hat{p}_{\mu }\hat{\Gamma}^{\mu \underline{0}}$ does not
commute with a part of the right generators, namely with $\hat{\mathbf{S}}%
\hat{\mathbf{B}}$, $\hat B^3_R$ and $p_{\mu }^{R}$. However, it
commutes with $\hat{\mathbf{S}}^{2}-\hat{\mathbf{B}}^{2}$ and with generator 
$S^3_R$ of $SU(2)_{int}$, that is why particles described by the
equation (\ref{26}), may have certain values $j_{1}+j_{2}$ and $S^3_R$.
And if the parity operator connects only states with different signs of $%
j_{1}-j_{2}$, then the operator $\hat{p}_{\mu }\hat{\Gamma}^{\mu }$ connects
states with all possible values $j_{1}$ and $j_{2}$ for a fixed sum $%
j_{1}+j_{2}$.

Thus, the invariant subspaces of the operators $\hat{\Gamma}^{\mu \underline{%
0}}$ are subspaces of polynomials of degree $2(j_{1}+j_{2})$ by $z$ with
fixed values $S^3_R$. In particular, such subspaces are the spaces of
polynomials $f(x,z,\cc\uz)$ and $f(x,{\underline z}{\vphantom{z}},\cc z)$ of
degree $2(j_{1}+j_{2})$, characterized by $S^3_R=\mp (j_{1}+j_{2})$.

Let's take a closer look at the case of spin 1/2. Linear functions from the
subspace $f(x,z,\cc\uz)$ depend on 4 variables $z^{\alpha },\cc\uz_{\dot{%
\alpha}}$, $S^3_R=-1/2$. For them, we have 2 equations of the form (\ref%
{26}) with different signs of the mass term. Each of these two equations
combines two irreps of the improper Poincar\'{e} group, characterized by the
same sign of $p_{0}\eta $. One can see that in the rest frame 
\begin{equation}
\varkappa ^{\underline{0}}=\varepsilon ms,\ \ \varepsilon =\eta \mathop{{\rm
sign}}p_{0}\mathop{{\rm sign}}S_{R}^{3}\ .
\end{equation}%
In total, for functions linear in orientation variables, we have 4
equations, their solutions in the rest frame at $p_{0}>0$ are given by
formulas (\ref{etam}). Although the particle and its antiparticle belong to
different subspaces with $S^3_R=\pm 1/2$, they are described by
equations with the same sign before the mass term.

Let us introduce into the consideration the generator of the group $U(1)$ of
phase transformations of orientation variables $z$, 
\begin{equation}
\hat{\Gamma}^{5}={\textstyle\frac{1}{2}}\left( z_{\;\;{\underline{a}}%
}^{\alpha }\partial _{\alpha }^{\;\;{\underline{a}}}-\cc z_{\;\;\dot{%
\underline{b}}}^{{\dot{\beta}}}\partial _{{\dot{\beta}}}^{\;\;\dot{%
\underline{b}}}\right) .  \label{chir}
\end{equation}%
These transformations, along with external and internal automorphisms of the
Poincar\'{e} group, can be considered as field symmetry transformations on
the Poincar\'{e} group; see Ref. \cite{GitSh09}. Polynomials of degree $n$
in $z,{\underline{z}}{\vphantom{z}}$ and of degree $m$ in $\cc z,\cc\uz$ are
eigenfunctions of the operator $\hat{\Gamma}^{5}$ with the eigenvalue $n-m$.
In the subspaces of the functions $f(x,z,\cc\uz)$ and $f(x,{\underline{z}}{%
\vphantom{z}},\cc z)$ its eigenvalues coincide with the chirality, defined
as $j_{1}-j_{2}$. The chirality operator $\Gamma ^{5}$ commutes with the
right and the left generators of the Poincar\'{e} group, but does not
commute with $\Gamma ^{\mu \underline{0}}$.

The matrix representation of the Dirac equation can be obtained from Eq. (%
\ref{26}) for the functions $f(x,z,\cc\uz)$ and $f(x,{\underline z}{%
\vphantom{z}},\cc z)$, which are polynomials of the first order in $z$.

For example, substituting linear in $z$ functions corresponding to the spin $%
1/2$ from the subspace $f(x,z,\cc\uz)$, 
\begin{equation}
f_{D}(x,z)=\chi _{\alpha }(x)z^{\alpha }+\cc\psi^{{\dot{\alpha}}}(x)%
\cc{\underline z}_{{\dot{\alpha}}}=Z_{D}\Psi _{D}(x),\ Z_{D}=(z^{\alpha }\;%
\cc{\underline z}_{{\dot{\alpha}}}),\ \Psi _{D}(x)={\binom{\chi _{\alpha }(x)%
}{\cc \psi^{{\dot{\alpha}}}(x)}},  \label{28}
\end{equation}%
into equation (\ref{26}) and making a comparison of the coefficients at $%
z^{\alpha }$ and $\cc{\underline z}_{\dot{\alpha}}$ in the left- and
right-hand sides, we obtain the Dirac equation 
\begin{equation}
(\hat{p}_{\mu }\gamma ^{\mu }-\varkappa )\Psi _{D}(x)=0,\quad  \label{29}
\end{equation}%
where $\varkappa =2\varkappa ^{\underline{0}}$. The action of the chirality
operator (\ref{chir}) amounts to the multiplication by $\gamma ^{5}/2$, $%
\gamma ^{5}=\mathrm{diag}\{\sigma ^{0},-\sigma ^{0}\}$,%
\begin{equation*}
\hat{\Gamma}^{5}f_{D}(x,z)=\frac{1}{2}Z_{D}\gamma ^{5}\Psi _{D}(x).
\end{equation*}

Similarly, considering polynomials of the second degree leads to the
10-component Duffin-Kemmer equations, and higher degrees to the Bhabha
equations.
Note that starting with 2nd-order polynomials in $z$, states of the form (\ref{etam1}) transforming according to
representations of $T_{[j\ 0]}$ or $T_{[0\ j]}$ are not, generally speaking, eigenfunctions of
$\hat{p}^{\mu }\hat{\Gamma}^{\mu\underline{0}}$, and in the rest frame, solutions of these equations contain components that transform according to other representations.

\section{Concluding remarks\label{S9}}

A classification of relativistic orientable objects and its connection to
the particle phenomenology is studied. This study is based on the
possibility to describe these objects by a scalar field on the Poincar\'{e}
group. Scalar functions $f(q)$, $q=(x,z)$, on the group depend on the
Minkowski space coordinates $x=\left( x^{\mu }\right) =\left( x^{0},\mathbf{x%
}\right) ,$ $\mathbf{x=\ }\left( x^{k}\right) ,$ as well as on the
orientation variables $z=\left( z_{\alpha \beta }\right) $ given by elements
of the matrix $Z\in SL(2,C)$. The functions have certain transformation
properties with respect to both groups $M(3,1)_{ext}$ and $M(3,1)_{int}$,
see Sec. \ref{S1}. We classify them with the help of maximal sets of
commuting operators constructed from generators of the groups $M(3,1)_{ext}$
and $M(3,1)_{int}$.

We suppose that different particles correspond to states described by
functions $f(x,z)$ that are not connected by transformations from $%
M(3,1)_{ext}$ (i.e., by a change of space-fixed reference frame). Such
states are classified by eigenvalues of Casimir operators and some functions
of the right generators of the group (i.e., generators of the group $%
M(3,1)_{int}$).

This way of the classification, applied to linear and quadratic functions of 
$z$ allows one to identify some of them with known elementary particles of
spin $0$,$\frac{1}{2}$, and $1$. Below, we list results of such an
identification (the enumeration is given by values of spin and chirality in
the massless limit).

\textbf{Spin 1/2, chirality $\boldsymbol{\pm 1/2}$.} Such particles appear
in quadruplets, but not in particle/antiparticle pairs.

In the massless case, we have 4 states and 4 particles -- two $SL(2,C)_{int}$
doublets of chirality $\lambda =-1/2$ (particles) and $\lambda =-1/2$ (their
antiparticles), interconnected by both space reflection and charge
conjugation. The particle and the antiparticle differ in charge signs $%
S^3_R$, each of them appear in one chiral state only. When reduced to $%
SU(2)_{int}\subset SL(2,C)_{int}$, these two-dimensional representations
remain irreducible.

In the real world, the fours fundamental fermions (in massless limit) have
identical properties, for example, $(e_{L}^{-},\nu _{L})$ and $(e_{R}^{+},%
\tilde{\nu}_{R})$ or $(d_{L},u_{L})$ and $(\tilde{d}_{R},\tilde{u}_{R})$.

In the massive case, we have 8 states and 4 particles, differing in sign $%
S^3_R$ and the parity ($\eta =1$ corresponds to particles and $\eta =-1$
corresponds to antiparticles). In the rest frame, states of one particle
differ by two spin projections $S_{3}=\pm 1/2$. The fours $(e^{-},\nu)$ and $%
(e^{+},\tilde{\nu})$ or $(d,u)$ and $(\tilde{d},\tilde{u})$ of fundamental
fermions have analogous properties, excluding the equality of masses.

\textbf{Spin 1, chirality $\boldsymbol{\pm 1}$.}

In the massless case, there appear 6 states and 6 particles: two $%
SO(3,C)_{int}\sim SL(2,C)_{int}$ triplets of chirality $\lambda =-1$
(particles) and $\lambda =1$ (antiparticles), interconnected by both space
reflection and charge conjugation. Components of the triplets differ in
charge $S^3_R=0,\pm 1$, each particle is in only one chiral state. When
reduced to $SU(2)_{int}\subset SL(2,C)_{int}$, these three-dimensional
representations remain irreducible.

In the massive case, we have 18 states and 6 particles, differing in quantum
numbers $S^3_R=0,\pm 1$ and in the parity, $\eta =1$ (three particles)
and $\eta =-1$ (three antiparticles). In the rest frame, states of one
particle differ by three projections of spin, $S_{3}=0,\pm 1$.

These two triples differ only in signs of chirality ($m=0$) or parity ($%
m\neq 0$). Therefore, if it turns out that the other charges of such
particles are the same, then they can be considered as different
polarization states of one particle. With this interpretation, we have one
triplet, each component of which has 2 ($m=0$) or 6 states ($m\neq 0$).

The triplet $(W^{-},W^{0},W^{+})$ of vector bosons (in massless limit) has identical properties
with the one considered above. 
In the massive case the triplet $(W^{-},Z^{0},W^{+})$ has similar properties, 
excluding the equality of masses.

\textbf{Spin 1, chirality $\boldsymbol{0}$.}

In the case $m\ne 0$, we have $SO(3,1)_{int}\sim SL(2,C)_{int}$ quadruplet
of particles, containing two truly neutral particles ($\eta =\pm 1$, 
$S^3_R=0$) and two charged particles ($\eta =1$, $S^3_R=\pm 1$). 
Each particle from the quadruplet can be in 3 states, which differ in the 
rest frame by three projections of the spin $S_{3}=0,\pm 1$. In the 
ultra-relativistic limit, remains only one state with zero projection of the 
spin on the direction of motion.

The case $S^3_R=0$ corresponds to a truly neutral hypothetical particle,
which is called the notoph in the literature.

\textbf{Chirality $\boldsymbol{0}$.} In the massless case $SO(3,1)_{int}\sim
SL(2,C)_{int}$ quadruplet of particles contain two truly neutral particles ($%
S^3_R=0$), differing by the values of $B_{3}^{R}$, and two charged
particles ($S^3_R=\pm 1$). Each particle from the quadruplet has only one
chiral state with zero projection of the spin on the direction of motion.

\textbf{Spin 0.} Scalar particles are represented by a singlet (independent
of $z$ functions $f(x)$) and a quadruplet (2nd order polynomials in $z$).
The quadruplet consists of two truly neutral particles ($\eta =\pm 1$, $%
S^3_R=0$) and two charged particles ($\eta =-1$, $S^3_R=\pm 1$). The
characteristics of the Higgs boson correspond to both the singlet and to the
component of the quadruplet with $\eta =1$.

This list exhausts orientable objects described by polynomials of the 0, 1st
and 2nd orders in orientation variables.

Among the states discussed above there exist states with the same $S_{R}^{3}$
that differ only by signs of the chirality (or the parity for $m\neq 0$).
The question is: to consider these states describing different particles, or
they are different polarization states of one and the same particle. If we
start only from the representation theory of the Poincar\'{e} group, we do
not get any answer to such a question. But, using the empirical formula (\ref%
{s3rLBQ}), we see that in the case of spin 1/2, these states differ in the
electric charge and, therefore, are states of different particles (electron
and neutrino). At the same time, in the case of the spin 1, Eq. (\ref{s3rLBQ}%
) does not prohibit interpreting them as different polarization states of
one and the same particle (photon, $W$ bosons).

When considering relativistic wave equations and particles described by
them, only 8 out of 10 from the maximal set of commuting operators
of the Poincar\'{e} group are usually used (6 left generators and 2 of 4
functions of the right generators, the latter define the irrep of the 
Lorentz group $SL(2,C)_{int}$). Here, for the classification of the 
orientable objects we have used the all 10 operators from the maximal set.

We stress that the results of developed classification, do not contradict
the phenomenology of elementary particles, and, moreover, in some cases,
present the corresponding group-theoretic explanation. Besides, the
relationship of the spectra of the left and right generators allows us to
give a possible explanation of the relationship between spin and the
internal quantum numbers (charges) of particles, expressed in the positivity
of $R$-parity.

In addition, it seems important to note that the composition and properties
of particles predicted by the presented classification follow exclusively
from the properties of the space-time and are, in a sense, model independent.

\section*{Appendix. Generators of the Lorentz group in the
space of functions $f\left( z\right)$\label{S10}}

The Lorentz group $SL(2,C)$ is a group of complex $2\times 2$ matrices with a unit determinant,
\begin{equation}
\label{SL2Cmarix}
Z=\left(\begin{array}{cc} z^1_{\;\,\underline{1}} & z^1_{\;\,\underline{2}} \\ z^2_{\;\,\underline{1}} & z^2_{\;\,\underline{2}}
 \end{array}\right) =
\left( \begin{array}{cc}
 z^1 & {\underline z}^1 \\
 z^2 & {\underline z}^2 \\
\end{array} \right),\quad
\det Z= z^1 {\underline z}^2 - z^2 {\underline z}^1 =1.
\end{equation}
For the sake of brevity, we have used the notation that we applied in \cite{GitSh09,GitSh01} and
$z^{\alpha }=z_{\;\;\underline{1}}^{\alpha }$, $\cc z_{\dot{\alpha}}=\cc z_{\dot{\alpha}}^{\;\;\dot{\underline{2}}}$,
${\underline{z}}^{\alpha }=z_{\;\;\underline{2}}^{\alpha }$,
$\cc {\underline z}_{\dot{\alpha}}=\cc z_{\dot{\alpha}}^{\;\;\dot{\underline{1}}}$.
In consequence of the unimodularity of $2\times 2$ matrices $Z$ there
exist invariant antisymmetric tensors
$\varepsilon ^{\alpha \beta}=-\varepsilon ^{\beta \alpha }$,
$\varepsilon ^{{\dot{\alpha}}{\dot{\beta}}}=-\varepsilon ^{{\dot{\beta}}{\dot{\alpha}}}$,
$\varepsilon ^{12}=\varepsilon ^{{\dot{1}}{\dot{2}}}=1$,
$\varepsilon _{12}=\varepsilon _{{\dot{1}}{\dot{2}}}=-1$.
Now spinor indices are lowered and raised according to the rules
\begin{equation}
\label{ind1}
z_{\alpha }=\varepsilon _{\alpha \beta }z^{\beta },\quad
z^{\alpha }=\varepsilon ^{\alpha \beta }z_{\beta },
\end{equation}
and, in particular
\begin{equation}
\label{ind2}
\cc Z=\left( \begin{array}{cc}
 \cc z^{\dot 1}_{\;\;\dot{\underline{1}}} & \cc z^{\dot 1}_{\;\;\dot{\underline{2}}} \\
 \cc z^{\dot 2}_{\;\;\dot{\underline{1}}} & \cc z^{\dot 2}_{\;\;\dot{\underline{2}}} \\
\end{array} \right) =
\left( \begin{array}{rr}
 -\cc z_{\dot 2}^{\;\;\dot{\underline{2}}} & \cc z_{\dot 2}^{\;\;\dot{\underline{1}}} \\
 \cc z_{\dot 1}^{\;\;\dot{\underline{2}}} & -\cc z_{\dot 1}^{\;\;\dot{\underline{1}}} \\
\end{array} \right) =
\left( \begin{array}{rr}
 -\cc z_{\dot 2} & \cc {\underline z}_{\dot 2} \\
 \cc z_{\dot 1} & -\cc {\underline z}_{\dot 1} \\
\end{array} \right).
\end{equation}

Besides the four-dimensional vector notation for spin operators,
it is also convenient to use a three-dimensional notation: $\hat
S^k =\frac 12\epsilon_{ijk} \hat S^{ij}$, $\hat B^k =\hat S_{0k}$.
In the space of functions on the group $f(z)$
(functions of the elements of the matrix $SL(2,C)$(\ref{SL2Cmarix}) and their complex conjugates)
a direct calculation yields for left and right generators \cite{GitSh01}:
\begin{eqnarray}
\label{SL} 
&&\hat S^k=\frac 12 (z\sigma^k\partial _z -\cc
z\cc\sigma^k\partial _{\ccc z}\,)+... \; ,
\nonumber \\
&&\hat B^k=
\frac i2 (z\sigma^k\partial _z + \cc z\cc\sigma^k\partial _{\ccc z}\,)+... \; ,
\quad z=(z^1\; z^2), \quad
\partial_z=(\partial/\partial{z^1}\; \partial/\partial{z^2})^T ;
\\   
\label{SR}
&&\hat S^k_R=-\frac 12 (\chi\cc\sigma^k\partial_\chi
-\cc\chi\sigma^k\partial _{\ccc \chi}\,)+... \; ,
\nonumber \\
&&\hat B^k_R=
-\frac i2 (\chi\cc\sigma^k\partial_\chi + \cc\chi\sigma^k\partial _{\ccc \chi}\,)+... \; ,
\quad \chi=(z^1\;\uz^1), \quad
\partial_\chi=(\partial/\partial{z^1}\; \partial/\partial{\uz^1})^T .
\end{eqnarray}
The dots here stand for analogous expressions obtained by
the change $z\to z'=(\uz^1\;\uz^2)$, $\chi\to \chi'=(z^2\; \uz^2)$.
In particular,
\begin{equation}
\label{S3B3}
\hat S^3_R= \frac 12 (-z\partial _z + \uz\partial _\uz + \cc z\partial _{\ccc z} - \cc \uz\partial _{\ccc \uz}\,) \; ,\qquad
\hat B^3_R= \frac i2 (-z\partial _z + \uz\partial _\uz - \cc z\partial _{\ccc z} + \cc \uz\partial _{\ccc \uz}\,) \; .
\end{equation}

It is known that from $\hat S^k$ and $\hat B^k$ one can construct
such linear combinations $\hat M^k$ and $\barM^k$,
\begin{eqnarray}
&&\hat M^k=\frac 12(\hat S^k-i\hat B^k)=z\sigma^k\partial _z + \uz\sigma^k\partial _\uz, \quad
\hat M_+=z^1\partial /\partial{z^2}, \quad
\hat M_-=z^2\partial /\partial{z^1},
\nonumber \\
&&\barM^k=-\frac 12(\hat S^k+i\hat B^k)=\cc z\cc\sigma^k\partial _{\ccc z} + \cc\uz\cc\sigma^k\partial _{\ccc\uz}\,, \quad
\barM_+=\cc\uz^{\dot 1}\partial /\partial{\cc\uz^{\dot 2}}, \quad
\barM_-=\cc\uz^{\dot 2}\partial /\partial{\cc\uz^{\dot 1}},
\label{MNgen}
\end{eqnarray}
that $[\hat M^i,\barM^k]=0$; in addition, for unitary representations
of the Lorentz group, as it follows from the condition
$\hat S^{k\dagger} =\hat S^k$, $\hat B^{k\dagger} =\hat B^k$,
the relation $\hat M^{k\dagger} =\barM^k$ must be fulfilled
(for finite-dimensional non-unitary representations
$\hat S^{k\dagger} =\hat S^k$,
$\hat B^{k\dagger} =-\hat B^k$ and $\hat M^{k\dagger} =-\barM^k$).

Taking into account the fact that the operators $\hat M^k$ and $\barM^k$
satisfy the commutation relations of the algebra
$su(2)$, we find the following relations for the spectra of the Casimir operators
of the Lorentz subgroup:
\begin{eqnarray}
&&\hat {\mathbf S}^2 -\hat {\mathbf B}^2 = 2(\hat {\mathbf M}^2+\barMM)
  =2(j_1(j_1+1)+j_2(j_2+1))=2(j_1+j_2+1)(j_1+j_2), \quad
\nonumber \\
&&\hat {\mathbf S}\hat {\mathbf B} = -i(\hat {\mathbf M}^2-\barMM)
  =-i\left( j_1(j_1+1)-j_2(j_2+1)\right) =-i(j_1+j_2+1)(j_1-j_2).
\label{SL2Ccas}
\end{eqnarray}
That is, the irreps of the Lorentz group $SL(2,C)$ are labeled
by a pair of numbers $[j_1,j_2]$.

The difference $j_1-j_2$ (the difference between the number of
dotted and undotted indices) can also be obtained as an eigenvalue
of the chirality operator $\hat \Gamma^5$ (\ref{chir}).

For finite-dimensional irreps of the group $SL(2,C)$, the formula of reduction to
the compact $SU(2)$-subgroup have the form 
\begin{equation}  
\label{SL2Cred}
T_{[j_1,j_2]}=\sum_{j=|j_1-j_2|}^{j_1+j_2}T_j. 
\end{equation}


\end{document}